\documentclass[twocolumn,twocolappendix]{aastex701} 

\usepackage{graphicx}
\usepackage{xcolor}
\usepackage[dvipsnames]{xcolor}
\usepackage{txfonts}
\usepackage{float}

\begin{document} 

\title{The JWST Early Release Science Program for Direct Observations of Exoplanetary Systems.\\
VII. Molecular Mapping Performance with JWST/MIRI MRS: VHS\,1256\,b as a case study}

\author[orcid=0000-0002-2918-8479,sname='Mathilde Mâlin']{Mathilde Mâlin}
\affiliation{Department of Physics \& Astronomy, Johns Hopkins University, 3400 N. Charles Street, Baltimore, MD 21218, USA}
\affiliation{Space Telescope Science Institute, 3700 San Martin Drive, Baltimore, MD 21218, USA}
\affiliation{LIRA, Observatoire de Paris, Univ. PSL, CNRS, Sorbonne Universit{\'e}, Univ. Paris Diderot, Sorbonne Paris Cit{\'e}, 5 place Jules Janssen, 92195 Meudon, France}
\email[show]{mmalin@stsci.edu}

\author[orcid=0000-0001-9353-2724,sname='Anthony Boccaletti']{Anthony Boccaletti}
\affiliation{LIRA, Observatoire de Paris, Univ. PSL, CNRS, Sorbonne Universit{\'e}, Univ. Paris Diderot, Sorbonne Paris Cit{\'e}, 5 place Jules Janssen, 92195 Meudon, France}
\email[]{}

\author[orcid=0000-0001-9353-2724,sname='Benjamin Charnay']{Benjamin Charnay}
\affiliation{LIRA, Observatoire de Paris, Univ. PSL, CNRS, Sorbonne Universit{\'e}, Univ. Paris Diderot, Sorbonne Paris Cit{\'e}, 5 place Jules Janssen, 92195 Meudon, France}
\affiliation{Laboratoire d’Astrophysique de Bordeaux, Univ. Bordeaux, CNRS, B18N, allée Geoffroy Saint-Hilaire, 33615 Pessac, France}
\email[]{}

\author[orcid=0000-0003-3818-408X,sname='Laurent Pueyo']{Laurent Pueyo}
\affiliation{Space Telescope Science Institute, 3700 San Martin Drive, Baltimore, MD 21218, USA}
\email[]{}

\author[orcid=0000-0002-9799-2303,sname='Alexis Bidot']{Alexis Bidot}
\affiliation{Space Telescope Science Institute, 3700 San Martin Drive, Baltimore, MD 21218, USA}
\email[]{}

\author[orcid=0000-0001-8718-3732,sname='Polychronis Patapis']{Polychronis Patapis}
\affiliation{ETH Zürich, Institute for Particle Physics and Astrophysics, Wolfgang-Pauli-Str. 27, 8093 Zürich, Switzerland}
\email[]{}

\author[orcid=0000-0001-8074-2562,sname='Sasha Hinkley']{Sasha Hinkley}
\affiliation{University of Exeter, Astrophysics Group, Physics Building, Stocker Road, Exeter, EX4 4QL, UK}
\email[]{}

\author[orcid=0009-0006-5349-5192,sname='Simon Petrus']{Simon Petrus}
\affiliation{NASA Goddard Space Flight Center, Greenbelt, MD 20771, USA}
\email[]{}

\author[0000-0001-8818-1544]{Niall Whiteford}
\affiliation{Department of Astrophysics, American Museum of Natural History, Central Park West at 79th Street, New York, NY 10034, USA}
\email[]{}

\author[orcid=0000-0002-3191-8151,sname='Marshall Perrin']{Marshall Perrin}
\affiliation{Space Telescope Science Institute, 3700 San Martin Drive, Baltimore, MD 21218, USA}
\email[]{}

\author[0000-0003-4614-7035]{Beth A. Biller}
\affiliation{Scottish Universities Physics Alliance, Institute for Astronomy, University of Edinburgh, Blackford Hill, Edinburgh EH9 3HJ, UK}
\affiliation{Centre for Exoplanet Science, University of Edinburgh, Edinburgh EH9 3HJ, UK}
\email[]{}

\author[orcid=0000-0001-7255-3251,sname='Gabriele Cugno']{Gabriele Cugno}
\affiliation{Department of Astrophysics, University of Zürich, Winterthurerstrasse 190, 8057 Zürich, Switzerland}
\email[]{}

\author[0000-0002-7405-3119]{Thayne Currie}
\affiliation{Department of Physics and Astronomy, University of Texas at San Antonio,  1 UTSA Circle, San Antonio, TX USA}
\affiliation{National Astronomical Observatory of Japan, Subaru Telescope, 650 N. Aohoku Pl, Hilo, HI USA}
\email[]{}

\author[orcid=0000-0002-3729-2663]{Camilla Danielski}
\affiliation{Departament d'Astronomia i Astrof\'{i}sica, Universitat de Val\`{e}ncia, Av. Vicent Andrés Estellés 19, 46100, Burjassot, Spain}
\affiliation{INAF – Osservatorio Astrofisico di Arcetri, Largo E. Fermi 5, 50125 Firenze, Italy}
\email[]{}

\author[0000-0002-1493-300X]{Thomas Henning}
\affiliation{Max-Planck-Institut f\"ur Astronomie, K\"onigstuhl 17, 69117 Heidelberg, Germany}
\email[]{}

\author[0000-0002-9803-8255]{Kielan K. W. Hoch}
\affiliation{Space Telescope Science Institute, 3700 San Martin Drive, Baltimore, MD 21218, USA}
\email[]{}

\author[orcid=0000-0001-8345-593X]{Markus Janson}
\affiliation{Department of Astronomy, Stockholm University, AlbaNova University Center, 10691 Stockholm, Sweden}
\email[]{}

\author[0000-0003-2769-0438]{Jens Kammerer}
\affiliation{European Southern Observatory, Karl-Schwarzschild-Str. 2, 85748, Garching, Germany}
\email[]{}

\author[orcid=0000-0003-0593-1560]{Elisabeth C Matthews}
\affiliation{Max-Planck-Institut f\"ur Astronomie, K\"onigstuhl 17, 69117 Heidelberg, Germany}
\email[]{} 

\author[orcid=0000-0002-9792-3121]{Evert Nasedkin}
\affiliation{School of Physics, Trinity College Dublin, University of Dublin, Dublin 2, Ireland}
\email[]{}

\author[0000-0002-6217-6867]{Paulina Palma-Bifani}
\affiliation{LIRA, Observatoire de Paris, Univ. PSL, CNRS, Sorbonne Universit{\'e}, Univ. Paris Diderot, Sorbonne Paris Cit{\'e}, 5 place Jules Janssen, 92195 Meudon, France}
\email[]{}

\author[0000-0002-4388-6417]{Isabel Rebollido}
\affiliation{Centro de Astrobiolog\'ia (CAB CSIC-INTA) ESAC Campus Camino Bajo del Castillo, s/n, Villanueva de la Cañada, 28692, Madrid, Spain}
\email[]{}

\author[0000-0001-9992-4067]{Matthias Samland}
\affiliation{Max-Planck-Institut f\"ur Astronomie, K\"onigstuhl 17, 69117 Heidelberg, Germany}
\email[]{}

\author[0000-0001-6098-3924]{Andrew Skemer} 
\affiliation{University of California Santa Cruz, Santa Cruz, CA, USA}
\email[]{}

\author[0000-0003-0454-3718]{Jordan M. Stone}
\affiliation{Naval Research Laboratory, Remote Sensing Division, 4555 Overlook Ave SW, Washington, DC 20375 USA}
\email[]{}

\author[0000-0002-2011-4924]{Genaro Suárez}
\affiliation{Department of Astrophysics, American Museum of Natural History, Central Park West at 79th Street, New York, NY 10034, USA}
\email[]{}

\author[0000-0002-9962-132X]{Ben J. Sutlieff}
\affiliation{Scottish Universities Physics Alliance, Institute for Astronomy, University of Edinburgh, Blackford Hill, Edinburgh EH9 3HJ, UK}
\affiliation{Centre for Exoplanet Science, University of Edinburgh, Edinburgh EH9 3HJ, UK}
\email[]{}

\author[orcid=0000-0002-6510-0681,sname='Motohide Tamura']{Motohide Tamura}
\affiliation{The University of Tokyo, Hongo, Tokyo 113-0033, Japan}
\affiliation{Astrobiology Center, Mitaka, Tokyo 181-8588, Japan}
\affiliation{The University of Osaka, Suita, Osaka 565-0871, Japan}
\email[]{}

\author[0000-0002-9807-5435]{Christopher A. Theissen}
\affiliation{Department of Astronomy \& Astrophysics, University of California San Diego, UCSD Mail Code 0424, 9500 Gilman Drive, La Jolla, CA 92093-0424, USA}
\email[]{}

\author[0000-0003-0489-1528]{Johanna M. Vos}
\affiliation{School of Physics, Trinity College Dublin, University of Dublin, Dublin 2, Ireland}
\affiliation{Department of Astrophysics, American Museum of Natural History, Central Park West at 79th Street, New York, NY 10034, USA}
\email[]{}

\author[0000-0002-3726-4881]{Zhoujian Zhang}
\affiliation{Department of Physics \& Astronomy, University of Rochester, Rochester, NY 14627, USA}
\email[]{}

\author[orcid=0000-0002-5903-8316]{Alice Zurlo}
\affiliation{Instituto de Estudios Astrof\'isicos, Facultad de Ingenier\'ia y Ciencias, Universidad Diego Portales, Av. Ej\'ercito Libertador 441, Santiago, Chile}
\affiliation{Millennium Nucleus on Young Exoplanets and their Moons (YEMS)}
\email[]{}

\collaboration{all}{and the Direct Observations of Exoplanetary Systems ERS collaboration}

\begin{abstract}
VHS\,1256\,b was the first planetary-mass companion to be observed with the James Webb Space Telescope's Mid-Infrared Instrument (JWST/MIRI) using the Medium-Resolution Spectrometer (MRS).
The MRS provides high-quality integral-field spectral data in the mid-infrared (IR) wavelengths from 4.9 to 18 $\mu$m.
This dataset serves as a testbed for applying cross-correlation techniques to characterize exoplanet atmospheres.
We implement the so-called molecular mapping approach, which consists of performing a spectral cross-correlation between each spectral pixel and atmospheric model templates.
We compare these results with those obtained from cross-correlation of the extracted spectrum.
Using a self-consistent \texttt{Exo-REM} atmospheric model grid, we constrain the temperature, surface gravity, C/O ratio, and metallicity, finding values consistent with those obtained from other analysis methods.
We detect CO (S/N $\sim$ 25) and H$_2$O (S/N $\sim$ 76), with tentative detections of NH$_3$ and CH$_4$ (S/N $\sim$ 3).
We test cross-correlation to measure trace-species abundances and isotopic ratios. 
We measure a volume mixing ratio of $[\rm NH_3] =-5.73^{+0.15}_{-0.14}$ and an isotopic ratio $^{12}\mathrm{C}/^{13}\mathrm{C}=77.8^{+13}_{-10}$, both consistent with free-chemistry retrievals.
The derived NH$_3$ volume mixing ratio, combined with the measured temperature and radius, is consistent with VHS\,1256\,b having a mass above the deuterium-burning limit.
These results demonstrate the diagnostic power of mid-IR spectroscopy and highlight cross-correlation as a robust method for characterizing directly imaged exoplanets, even in future higher-contrast regimes where spectral extraction becomes challenging.
Future MIRI MRS observations across a wider range of temperatures and masses will further expand our understanding of planetary atmospheric chemistry.

\end{abstract}


\section{Introduction}

The diversity of known imaged exoplanets presents many challenges to understand the formation and evolution of planetary systems
\citep{currie_direct_2023}.
Their evolution can be probed through atmospheric metallicity and abundance measurements, including elemental ratios such as C/O and, more recently, isotopic ratios like $^{12}$CO/$^{13}$CO and $^{14}$N/$^{15}$N \citep[e.g.][]{konopacky_detection_2013, ruffio_deep_2021, hoch_assessing_2023, palma-bifani_peering_2023, zhang_13co-rich_2021, gandhi_jwst_2023, barrado_15nh3_2023, kuhnle_water_2025}.
Elemental and isotopic abundance ratios trace the chemistry of the protoplanetary disk and thus provide key insights into planet formation pathways, with radial variations in disk C/O ratios illustrating how chemical gradients shape planetary compositions \citep[e.g.][]{oberg_effects_2011, oberg_astrochemistry_2021, turrini_tracing_2021, nomura_isotopic_2022}.
However, inferring these ratios requires accurate molecular abundance measurements, which are often challenging for exoplanet atmospheres observed over limited spectral ranges, and the C/O ratio itself is not a fixed property but evolves with disk chemistry over time, further complicating its interpretation \citep{molliere_interpreting_2022}.

Young brown dwarfs span a similar range of temperatures, surface gravities, and masses as self-luminous super-Jupiter exoplanets, implying comparable atmospheric physics and chemistry, 
while their higher S/N, higher spectral resolution, and broader wavelength coverage enable more precise C/O measurements.
Together with free-floating planetary-mass analogs, they therefore serve as powerful laboratories for giant-planet atmosphere studies, benefiting from lower stellar contrast and consequently higher photometric accuracy \citep{petrus_x-shyne_2025, faherty_signatures_2014, bonnefoy_library_2014}.
Observations as part of the James Webb Space Telescope (JWST) Early Release Science (ERS) Program for Direct Observations of Exoplanetary Systems \citep[ERS 1386, detailed in ][]{hinkley_jwst_2022} include high signal-to-noise ratio ($S/N$), medium-resolution spectra of a companion lying at the boundary between planetary and brown-dwarf masses, spanning the near- to mid-infrared (IR) wavelengths.

\subsection{VHS\,1256: a Rosetta Stone for giant exoplanet atmospheric characterization}
The system VHS\,J125601.92-125723.9, (hereafter VHS\,1256), is a hierarchical system with a tight equal-mass M7.5 binary \citep[separated from 0.1$''$,][]{stone_adaptive_2016} and the circumbinary companion VHS\,1256\,b \citep{gauza_discovery_2015} is at a projected physical separation of 179~$\pm$~9~au (8.06~$\pm$~0.03$''$). 
It is located at 21.15~$\pm$~0.22~pc
\citep{gaia_collaboration_gaia_2023}.
It has a derived age of 140~$\pm$~20~Myr, and the companion's mass was inferred using its luminosity value \citep{dupuy_masses_2022}. 
The inferred masses form a bimodal distribution, with one solution at $12 \pm 0.1~M_{\rm Jup}$ and another at $16 \pm 1~M_{\rm Jup}$, placing the object close to the deuterium-fusion boundary \citep{dupuy_masses_2022, miles_jwst_2023}.
Its low surface gravity consistent with its young age, and atmospheric reddening confirms the presence of cloud layers \citep{stephens_08-145_2009}.
Its youth, temperature, low gravity, and mass make it an ideal laboratory for testing post-processing methods for characterizing young giant exoplanets, as these properties closely overlap with those of directly imaged exoplanets \citep{dupuy_distances_2013}.
Indeed, the spectroscopic characteristics and photometric colors of this object are similar to those of directly imaged exoplanets, such as the well-studied planets HR\,8799\,c, d and e \citep{faherty_population_2016}, with an estimated temperature ranging from 1100--1300\,K \citep{boccaletti_imaging_2024}, placing it at the boundary of the L--T transition.
Methane has been detected in low abundance in its atmosphere, suggesting non-equilibrium chemistry between CO and CH$_4$ \citep{miles_methane_2018}. 
Based on medium-resolution spectral analysis, the object exhibits super-solar metallicity and a super-solar $\mathrm{C/O}$ ratio.
The metallicity of the binary host stars is consistent with solar values \citep{guirado_radio_2018}.
Specifically, the metallicity was measured at $[\mathrm{M/H}] = 0.21 \pm 0.29$ \citep{petrus_x-shyne_2023}, and the $\mathrm{C/O}$ ratio was reported as either $> 0.63$ \citep{petrus_x-shyne_2023} or $0.59^{+0.28}_{-0.35}$ \citep{hoch_assessing_2023}.
These values suggest an enrichment in solids during its formation.

Analysis of the JWST Near InfraRed Spectrograph (NIRSpec) and the Mid InfraRed Instrument (MIRI) spectrum leads to a robust bolometric luminosity for VHS\,1256\,b of $\log(L_{\rm bol}/L_\odot) = -4.550 \pm 0.009$, with derived atmospheric parameters T$_{\rm eff}=1100$ K, $\log g = 4.5$, and a radius of 1.27 R$_{\rm Jup}$, based on \texttt{PICASO 3.0} models \citep{miles_jwst_2023, mukherjee_picaso_2023}.
Using the Bayesian framework \texttt{ForMoSA}, \citet{petrus_jwst_2024} tested the latest generation of self-consistent atmospheric models on the same spectrum.
The estimate of each parameter is significantly influenced by factors such as the wavelength range considered, the spectral resolution, the $S/N$, and the model chosen.
The observed parameter dispersion may be attributed to systematic incompleteness in the models, resulting from their difficulties in accurately replicating the complex atmospheric structure of VHS 1256 b. 
In particular, modeling their cloud structures and dust distributions remains highly challenging, and the state-of-the-art for such objects is advancing rapidly \citep{luna_empirically_2021, moran_neglected_2024}.
\cite{petrus_jwst_2024} concluded that despite the exceptional data quality, attaining robust estimates for chemical abundances [M/H] and $\mathrm{C/O}$, often employed as indicators of formation history, remain challenging to constrain, with estimates dispersed throughout the range explored by the model grids.
The measured metallicity and elemental abundance ratios are sensitive to assumptions about atmospheric structure, whether the atmosphere is cloudy, with homogeneous or patchy clouds, and to the choice of priors.
This remains true even in free-chemistry retrieval frameworks that allow for complex inhomogeneities such as patchy clouds and hot spots \citep{zhang_elemental_2025}. 
The JWST spectrum reveals a complex atmosphere impacted by both silicate clouds and non-equilibrium chemistry. 
The MIRI data exhibit a pronounced 10\,$\mu$m silicate absorption feature associated with silicate clouds \citep{miles_jwst_2023}. 
Notably, the depth of this feature exceeds that typically observed in objects of comparable effective temperature \citep{suarez_ultracool_2022}.
The strength of CH$_4$ and CO absorption features suggest relative abundances that deviate from thermochemical equilibrium in the photosphere, indicating that vertical mixing must be strong enough to drive transport-induced disequilibrium chemistry \citep{miles_methane_2018}. 

VHS\,1256\,b has also been previously identified as one of the most variable substellar objects, further evidence of a complex and heterogeneous atmospheric structure. 
The higher amplitude and increased frequency of variability in L-T transition objects has been attributed to changes in the vertical \citep{vos_patchy_2023, apai_hst_2013}, 
and longitudinal distribution of clouds \citep{nasedkin_jwst_2025}. 
The presence of both silicate clouds and non-equilibrium chemistry in VHS\,1256\,b is consistent with expectations for objects in the L-T transition temperature regime 
\citep{charnay_self-consistent_2018, saumon_evolution_2008}, 
but 1D radiative convective equilibrium forward models struggle to fully reproduce the details of observations in this complex regime.

More recently, a free-retrieval analysis using the \texttt{Brewster} framework favored a combination of patchy clouds of amorphous silicates with forsterite (Mg$_2$SiO$_4$) and enstatite (MgSiO$_3$) stoichiometries,
to reproduce the observed silicate absorption feature (Whiteford et al. subm).
The inferred cloud patchiness is consistent with the previously reported spectral variability \citep{zhou_spectral_2020, bowler_strong_2020}, 
and more recent work shows that the variability cannot be attributed to a single varying atmospheric property \citep{lueber_retrieved_2024}, suggesting that heterogeneous cloud coverage remains the most plausible origin. 
That study also retrieved atmospheric parameters, including molecular abundances of H$_2$O, CH$_4$, CO, CO$_2$, and NH$_3$, and highlighted the sensitivity of the retrieved quantities to the specific data sets and the relative $S/N$ ratios across different instruments.

Carbon and oxygen isotopologues have recently been measured using JWST/NIRSpec spectra in the 4.1--5.3\,$\mu$m range, which provide strong constraints on the isotopologues of CO \citep{gandhi_jwst_2023}. 
Specifically, the isotopic ratio $^{12}\mathrm{C}/^{13}\mathrm{C}$ is an interesting observable, as it varies between exoplanets and brown dwarfs \citep{zhang_13co-rich_2021} and also differs between the Solar System and the local interstellar medium \citep{milam_12_2005}. 
For VHS~1256\,b, the measured value of $^{12}\mathrm{C}/^{13}\mathrm{C} = 62^{+2}_{-2}$ \citep{gandhi_jwst_2023} lies between those typically found for planetary-mass companions and isolated brown dwarfs \citep{gonzalez_picos_eso_2024-1}.

\subsection{Molecular Mapping: a powerful tool to characterize exoplanets' atmospheres}

The first successful detection of molecules in the atmosphere of a directly imaged planet (specifically CO and H$_2$O for the planet HR 8799c) was achieved through cross-correlation with molecular spectral models, using the planet's medium-resolution spectrum after continuum subtraction \citep{konopacky_detection_2013}.
Building on earlier concepts, \citep[e.g.,][]{sparks_imaging_2002, thatte_very_2007}, the combinaison of high-contrast imaging with high-resolution spectroscopy has been proposed to greatly enhance planet detection through correlation techniques \citep{snellen_combining_2015}.
This combination underpins the co-alled molecular mapping method, which has been successfully applied to archival VLT/SINFONI observations of $\beta$ Pic b, leading to the detection of H$_2$O and CO in its atmosphere \citep{hoeijmakers_medium-resolution_2018}.
%
The method exploits the distinct spectral signatures of stellar and planetary sources, enabling their spatial and spectral disentanglement with an integral field spectrograph
\citep[e.g.,][]{petit_dit_de_la_roche_molecule_2018, ruffio_radial_2019,petrus_medium-resolution_2021, cugno_molecular_2021, kiefer_new_2024}.
Its potential has been studied for future instruments \citep{houlle_direct_2021, bidot_exoplanets_2023, landman_trade-offs_2023} and has shown promise for characterizing the atmospheres of Earth-like planets \citep{snellen_combining_2015, vaughan_behind_2024}.
%
Technique that uses cross correlation between a template theoretical spectrum and direct spectroscopic data, benefits greatly from only considering the continuum subtracted information content.
Indeed, in the presence of systematic noise stemming from diffracted starlight, 
high-pass filtering in the spectral dimension yields deeper sensitivities to faint companions 
than when seeking to also measure the pseudo-continuum \citep{ruffio_jwst-tst_2024}. 

The molecular mapping technique for MIRI MRS data \citep{patapis_direct_2022, malin_simulated_2023} was initially explored through the use of \texttt{MIRISim} simulations \citep{klaassen_mirisim_2020}.
Recent work has focused on predicting the effective detection limits of spectro-imaging data analyzed with molecular mapping
\citep{martos_combining_2025}.
The recent detection of molecular emission from circumplanetary disk gas with MIRI-MRS further highlights the power of cross-correlation techniques \citep{cugno_carbon-rich_2025, malin_jwst-tst_2025}.
Similar spectral template cross-correlation techniques have recently been applied to detect and characterise closely separated companions with NIRSpec.
\citep[e.g.][]{madurowicz_direct_2025, ruffio_jwst-tst_2024}.
In the mid-IR, the wavelength coverage of MIRI MRS further enhances this approach by enabling the disentanglement of chemical and cloud contributions,
highlighting the particular importance of this method at MRS wavelengths.

\subsection{Objectives of this study}
The exquisitely deep data of the VHS\,1256\,b planetary-mass companion with MIRI MRS represent an ideal case for an empirical study of the advantages and limits of the molecular mapping technique and a comparison with pre-flight simulations. 
By focusing on a favorable case in which stellar contamination is negligible, we can explore the detectability of a wide range of molecular species under ideal conditions. 
Any challenges encountered in this dataset are likely to be present in other targets with higher stellar noise and contrast.
%
%
This approach allows us to evaluate its ability to retrieve the fundamental bulk properties of the exoplanet even after continuum subtraction, and to explore the measurement of molecular abundances for trace species.
These are the main aspects this paper aims to address.
Sect.~\ref{sec:data_methods} describes the observations, data reduction, and the methodology used for the cross-correlation analysis.
Sect.~\ref{sec:mm_application} details the results of the molecular mapping analysis and its limitations.
This methodology will be directly applicable for future fainter closer-in companions.
The following sections aim at comparing our approach with traditional cross-correlation methods, in order to better understand how exoplanets can be characterized using the molecular mapping technique.
Sect.~\ref{sec:ccf_1d} presents the 1D cross-correlation analysis with the spectrum directly extracted from the data; and
Sect.~\ref{sec:measuring_abundance} introduces the effort to measure molecular abundances and isotopic ratios with cross-correlations in comparison to free-chemistry retrieval analysis.
Finally, Sect.~\ref{sec:discussion} and Sect.~\ref{sec:conclusion} present the discussion and the conclusion.

\section{Data and methods}
\label{sec:data_methods}
\subsection{Observations and data reduction}
\begin{figure*}[t]
    \centering
    \includegraphics[width=18cm]{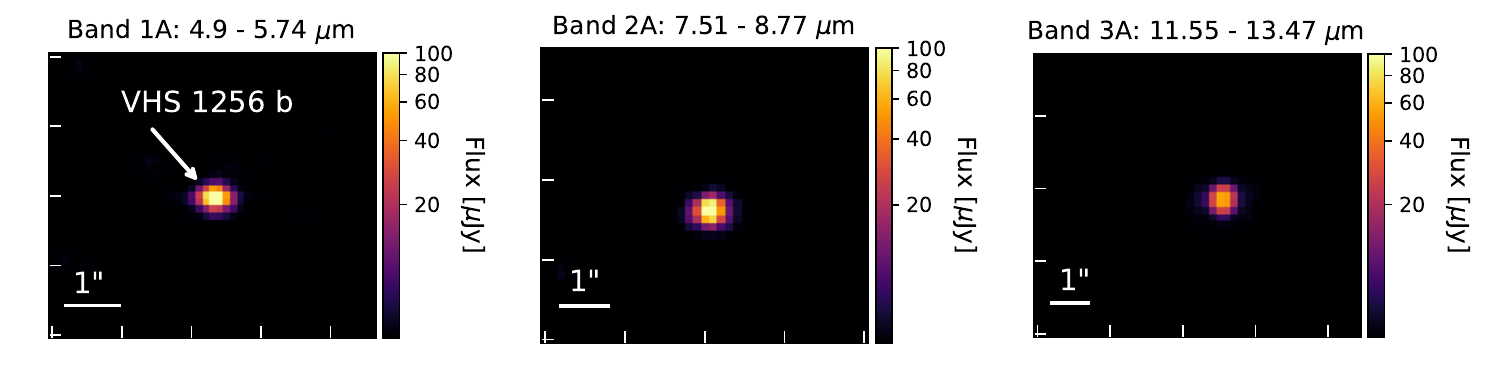}
    \caption{Median cubes in the band A of each MRS channel. VHS\,1256\,b is identified by the arrow. The color scale is the same for each band.}
    \label{fig:cubesA}
\end{figure*}
The Direct Observations of Exoplanetary Systems ERS\,1386 program includes the observations of the planetary-mass companion VHS\,1256\,b \citep{hinkley_jwst_2022}.
The data were observed 
on 2022-07-05 with both JWST integral field spectrographs: NIRSpec integral field unit (IFU) and MIRI MRS.
The extracted spectrum spans wavelengths from 1 to 20 $\mu$m with a resolution ranging from 1000 to 3700 \citep{miles_jwst_2023}. 
Our analysis focuses on the mid-IR part of this dataset obtained with the MRS.
The MRS observations were performed with 
142 groups and 1 integration.
The observations have been carried out with the 4-Point dither pattern to improve spatial sampling and mitigate bad pixels \citep{law_3d_2023}.
This results in a total exposure time of 
26.3 minutes.
%
Uncalibrated data are retrieved from the Mikulski Archive for Space Telescopes \citep[MAST,][]{marston_overview_2018} online data archive and reduced using the three stages of the \texttt{JWST} pipeline \citep{bushouse_jwst_2025} and default parameters. 
The data reduction method is similar to that detailed in \cite{miles_jwst_2023}, using the updated \texttt{JWST} pipeline version 1.16 and the Calibration References Data System CRDS version 1298.
Several improvements over the previous data reduction are available, including updated photometric calibrations and dark reference files, as well as improved cosmic ray shower correction.
The wavelength solution was improved and the aperture correction factors at each wavelength were also updated \citep{argyriou_jwst_2023}.
The cubes are reconstructed with the drizzle algorithm \citep{law_3d_2023} using IFU coordinates.

Even though the program contains sky background observations, they are unusable as they contain “showers” of cosmic rays, which appear as diffuse and extended structures on the detector \citep{miles_jwst_2023}. 
These structures differ from those expected from cosmic rays and, consequently, cannot be corrected by the \texttt{JWST} pipeline.
As a result, the background must be evaluated directly from the data.
The background contribution has little impact on the results of molecular mapping, as it is a low-frequency contribution in the spectral dimension.
However, since it begins to dominate at wavelengths longer than 10 $\mu$m, it is still preferable to subtract it from the data. 
The background is estimated together with detector effects, taking advantage of the 4-point dither observations.
At each position, we identified the pixel with the minimum flux across the four detector images from the dither positions. 
This allows us to reconstruct a model image of the detector effects, including the background, but without the signal of the object.
We subtract this background from each detector image before reconstructing the cubes.
This method appears to be the most effective for point-source objects.
It provides cleaner data cube for cross-correlation analysis and produces fewer correlation artifacts (see Sect.~\ref{sec:limitations}). 
A median image of band 1A, 2A, and 3A is shown in Fig.~\ref{fig:cubesA}, after subtracting the background contribution. 
The planetary-mass companion VHS\,1256\,b is clearly visible in the center of each image.
The molecular mapping approach described in Sect.~\ref{sec:mm_application} operates directly on these cubes to characterize the companion.

\subsection{Spectral extraction}
\begin{figure*}
    \centering
    \includegraphics[width=18cm]{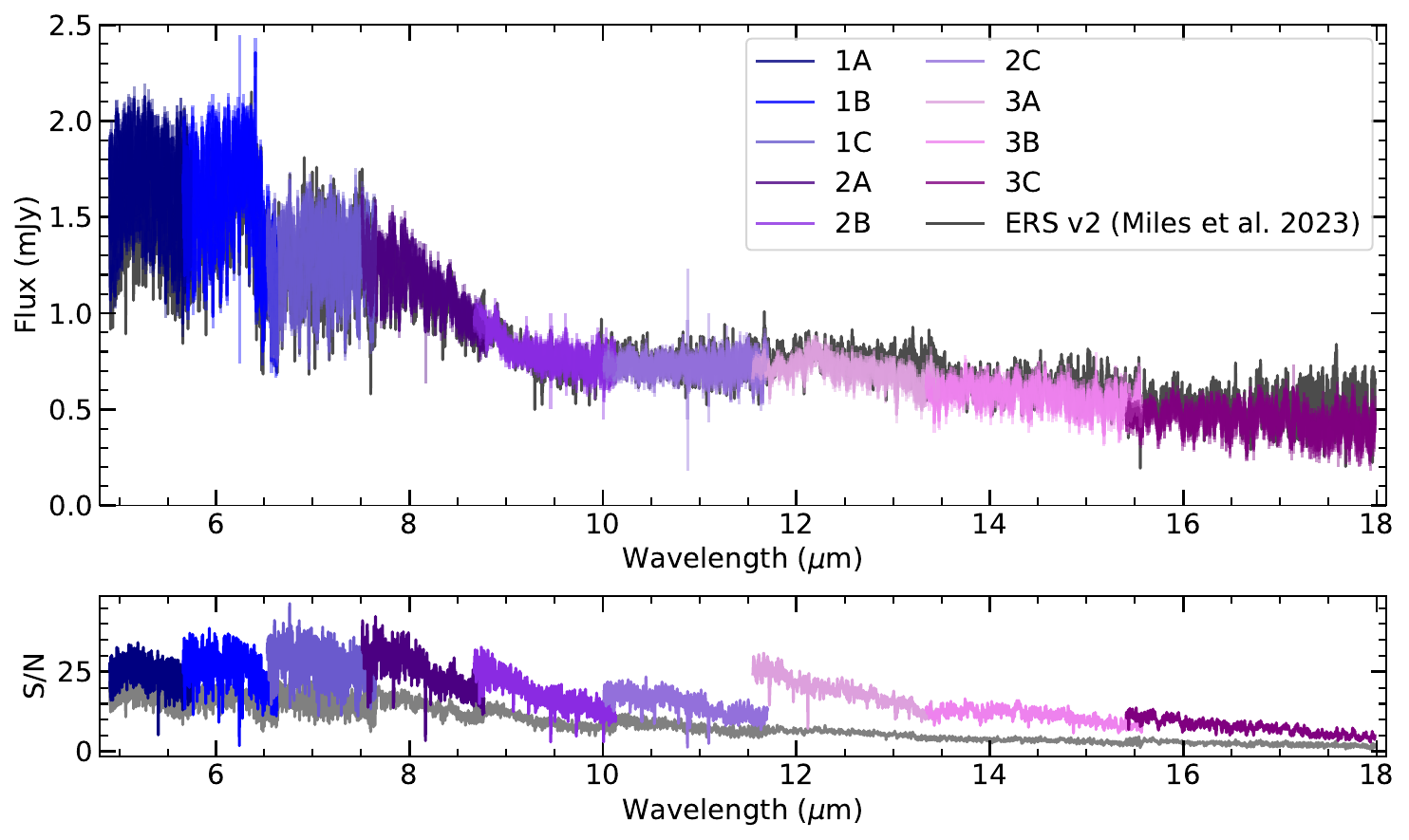}
    \caption{Updated spectra for VHS\,1256\,b in comparison to previously published spectra \citep{miles_jwst_2023} in grey.
    The $S/N$ per wavelength is indicated in the bottom subplot. }
    \label{fig:spectra}
\end{figure*}

The VHS 1256-1257 binary system lies outside the field of view and does not affect the detection of the companion, resulting in a high $S/N$ spectrum ($\sim$ 25 per wavelength, and decreasing to $\sim$ 10 at wavelengths longer than 14 $\mu$m).
This represents an ideal case for comparing the molecular mapping technique with a traditional cross-correlation approach.
The MIRI IFU sensitivity sharply decreases in channel 4 \citep{wells_mid-infrared_2015} and the thermal background emission increases,
this results in a weak detection at $\lambda$ $\gtrsim$ 18 $\mu$m.
Therefore, we limit our analysis to channels 1--3 (4.9--18 $\mu$m).
A 2D Gaussian is fitted to the median cube of each band to measure the exact position of the target.
The planet's signal is extracted at each wavelength in an aperture with a radius equivalent to the full width at half maximum of the point spread function (PSF).
Finally, we derived the flux density at each wavelength by applying the aperture correction from the reference files (\texttt{apcorr}).
The resulting spectrum is shown in Fig.~\ref{fig:spectra}
\footnote{The updated MRS spectrum can be accessed publicly at \href{https://doi.org/10.5281/zenodo.17858183}{zenodo.17858183}.}.

The \texttt{JWST} calibration pipeline provides an estimate of the flux density uncertainty at each wavelength.
These uncertainties are propagated through the pipeline but are generally underestimated, as it mainly accounts for photon noise and basic instrumental effects. 
Additional sources of uncertainty are not fully captured, including residual flat-field errors, imperfect background subtraction, spectral fringing, detector non-linearity, and time-dependent variations in the response. 
Consequently, 
we include an additional term in quadrature at each wavelength 
to obtain a more realistic estimate of the uncertainties. 
This term is derived from the standard deviation measured across a dozen apertures, placed at equal distances from the source and matched in size to the aperture used to extract the object’s flux.
This approach captures residual noise not accounted for by the pipeline and yields a more representative uncertainty for the aperture flux measurements.
While the error from the \texttt{JWST} pipeline remains dominant, this added contribution accounts for approximately 20--30\% of the total uncertainty.
This spectrum is consistent with the previously published results in \cite{miles_jwst_2023}, taking into account updates to the \texttt{JWST} pipeline and calibration files.
The uncertainties have been revised following the latest \texttt{JWST} calibrations, which provide improved spectral quality, and now incorporate the updated errors derived from our empirical estimates of the background noise. 
The improvement likely stems from better handling of straylight and cosmic rays in the latest \texttt{jwst} pipeline reduction, together with a cleaner background subtraction at longest wavelengths.

\subsection{\texttt{Exo-REM}}
\label{sec:pres_exorem}
Throughout the paper, we use the \texttt{Exo-REM} atmospheric model, a self-consistent 1D radiative-equilibrium model specifically developed for simulating the atmospheres and spectra of young giant exoplanets \citep{baudino_interpreting_2015,
charnay_self-consistent_2018, blain_1d_2021}.
\texttt{Exo-REM} was used to evaluate the performance of cross-correlation methods with MIRI MRS simulations \citep{malin_simulated_2023}, 
and has recently been shown to provide the best fit to the spectrum of VHS\,1256\,b \citep{petrus_jwst_2024}.
The model operates by solving radiative-convective equilibrium, assuming that the net flux is conservative across the pressure grid.
It includes disequilibrium chemistry, comparing chemical reaction timescales and vertical mixing using the parametrization of \cite{zahnle_methane_2014}, 
and considers clouds of iron and silicates (forsterite),
computed with simple microphysics \citep{charnay_self-consistent_2018}.
The sources of opacity include the H$_2$–H$_2$, H$_2$–He, H$_2$O–H$_2$O, and H$_2$O–air collision induced absorption and ro-vibrational bands from molecules (H$_2$O, CH$_4$, CO, CO$_2$, NH$_3$, PH$_3$, TiO, VO, H$_2$S, HCN, and FeH).
We use a pre-computed grid of atmospheric models, varying the temperature, metallicity, surface gravity, and the $\mathrm{C/O}$ ratio, as described in Table \ref{tab:table_grid_exorem}.
When looking for single molecules, as opposed to using the exoplanet's atmospheric model for cross-correlation, the molecular templates are calculated from the pressure–temperature profile at equilibrium and from previously calculated abundance profiles. 
The radiative transfer is computed again with all chemical species removed except the one considered.
The clouds are also removed, but collision-induced absorptions are included.
These spectra are shown in
Fig.~\ref{fig:ExoREM_molec_templates}.
The \texttt{Exo-REM} model spectra are degraded to the maximum resolution (i.e., R$_\lambda$ = 3700) of the MRS and are interpolated on the wavelength values of each MRS channel considered.
This study uses high-pass filtered data, and the same filtering is applied consistently to these atmospheric model spectra.

\begin{table}[h!]
\centering
\caption{Parameters of the \texttt{Exo-REM} grid.}
\begin{tabular}{c|  c}
\hline
\hline
Parameters & Bounds \\
\hline
T (K) & 400 -- 2000 (step: 50) \\
$\mathrm{log} g$ & 3.0 -- 5.0 (step: 0.5)\\
$\mathrm{C/O}$ & 0.1 -- 0.8  (step: 0.05) \\
Metallicity Z/Z$_\sun$ & 0.32 ; 1.0 ; 3.16 ; 10.0\\
\hline
\end{tabular}
\label{tab:table_grid_exorem}
\end{table}

\subsection{Cross-correlation processing and detection metrics}
\label{sec:cross_correlation}

Because VHS\,1256\,b is sufficiently bright and separated from host stars, its spectrum is not contaminated by diffracted starlight.
Cross-correlation between data and models can be measured in different ways.
(1) Molecular mapping: correlation applied directly to the data cube for each spaxel (i.e. spectral pixel) without extracting the spectrum of the object. 
This method provides 2D correlation maps, from which correlation values can be extracted at the planet’s central position.
Results are shown in Sect.~\ref{sec:mm_application}.
(2) 1D correlation: cross correlation applied to the extracted spectrum displayed in Fig.~\ref{fig:spectra}. 
This method provides the Cross Correlation Function (CCF) in 1D, and these results are shown in Sect.~\ref{sec:ccf_1d}.
It can be applied here because the stellar contribution is negligible, enabling the direct extraction of an uncontaminated spectrum.
This system provides an ideal test case to explore which atmospheric parameters can be constrained using only the high-frequency spectral variations (i.e., closely spaced molecular features), independent of the broader continuum in a directly imaged exoplanet spectrum.
%
%
%
We take advantage of this high $S/N$ dataset to compare both methods, exploring its limits and potential for future cases where extracting a planet's spectrum would be challenging due to contamination from the stellar signal.

First, the pseudo-continuum is removed from both the models and the data before performing cross-correlation.
To achieve this, we applied a Gaussian filter with $\sigma = 10$ (corresponding to a kernel of 10 wavelength points) to create a smoothed version of the spectrum, that is then subtracted.
This procedure has been shown to maximize detection in the simulated data \citep{malin_simulated_2023}.
We next define how $S/N$ is computed for each method, as these metrics will be discussed throughout the remainder of the paper.

\subsubsection{Signal-to-noise of detections}

\paragraph{Molecular mapping}
We apply the molecular mapping method exactly as developed and tested on simulated data \citep[see Sect.~3 in][]{malin_simulated_2023}\footnote{Publicly available tools for computing correlation maps are available at \href{https://github.com/mathildemalin/ExoCAT}{github.com/mathildemalin/ExoCAT}.}.
In each spaxel, low spectral frequencies are filtered out as described above, and the resulting high-frequency signal is correlated with a template.
The correlation value for each spaxel yields the 2D correlation map.
Many metrics have been used to estimate the level of detection on correlation maps. 
We adopt the $S/N$ measurement specifically developed for MIRI MRS data, which has been shown to provide more reliable estimates at low $S/N$ \cite{malin_simulated_2023}.
The signal of the companion is defined in an aperture centered on it, whose size depends experimentally on the mean azimuthal profile of the footprint of the correlation footprint.
We note that this correlation footprint is different from the standard concept of a PSF.
The noise is estimated as the standard deviation of the correlation values in the remaining spaxels, excluding the central object, using a 6-pixel-radius mask \citep[corresponding to the maximum extent of the footprint of the correlation pattern seen in simulations][]{malin_simulated_2023}.
This method is employed in the context of Sect.~\ref{sec:mm_application}.

Compared with the simulated data \citep{malin_simulated_2023},
we find that the $S/N$ in the correlation maps can be artificially increased when a few pixels at the planet’s position exhibit slightly higher correlation values, even in the absence of a true detection.
One way to assess that is to subtract a correlation map obtained with a featureless reference spectrum, meaning a model containing no molecular features
(as illustrated in Fig.~\ref{fig:artifacts}).
This test helps determine whether the measured correlation genuinely arises from molecular features or is instead dominated by residual noise structures.
This approach effectively traces the spatial distribution of noise and helps identify artifacts in low-$S/N$ regimes.

\paragraph{1D cross-correlation}
We compute the CCF over a range of velocity offsets between the extracted spectrum of VHS\,1256\,b and model template spectra after pseudo-continuum removal.
We define the $S/N$ as the peak of the CCF divided by the standard deviation of its wings ($\pm$ 500 to $\pm$ 3000 km/s).
The CCF wings can be strongly influenced by the template’s autocorrelation, especially for molecular spectra with regularly spaced features or harmonics, such as CO.
Therefore, we correct the CCF to remove the autocorrelation signal before measuring its noise level \citep{cugno_molecular_2021}.
We first compute the autocorrelation function (ACF) and subtract it from the CCF wings before measuring the $S/N$, preventing the molecular feature autocorrelation from inflating the noise estimate.
The application of this method is described in Sect.~\ref{sec:ccf_1d}.

\subsubsection{Statistical significance between two models}

We use the statistical framework described below to compare correlations between the data and various models and to quantify how strongly one model is favored over another.
This hypothesis testing 
in correlation space is not straightforward, as correlation values do not intrinsically follow a $\chi^2$ distribution.
We therefore adopt the formalism of \citet{zucker_cross-correlation_2003}, which maps the correlation coefficient to a log-likelihood function $\mathrm{log}(\mathcal{L})$ for hypothesis testing, such as: 
\begin{equation}
    \mathrm{log}(\mathcal{L}) = - \frac{N}{2}\mathrm{log}(1-C^2),\\
    \label{eq:cc-to-likelihood}
\end{equation}
$N$ is the number of bins in the spectrum and C the correlation value.
This approach still implies that our results follow a normal distribution, allowing for a consistent derivation of uncertainties.
Following Eq.~\ref{eq:cc-to-likelihood}, we compute $\log(\mathcal{L})$ for each model and derive $\Delta\chi^2$ using
$\Delta\chi^2 = -2\Delta\log(\mathcal{L})$.
For example, with one degree of freedom, the corresponding $\Delta\chi^2$ values are 0.99, 3.98, and 8.80 for the 1-, 2-, and 3-$\sigma$ confidence intervals, respectively (as used in Sect.\,\ref{sec:measuring_abundance}).
%
However, in practice, the data are not always well-behaved and may deviate from Gaussian statistics.
Although the noise component is additive and can be described by a $N$ dimensional normal distribution, the probability density function of the correlation with a template is typically non-Gaussian and skewed toward unity.
Moreover, its variance does depend on its mean: for instance, the variance of a correlation histogram is much smaller when most values are close to unity than when they are close to zero.
As a consequence, our method may yield slightly optimistic uncertainty estimates.
Nonetheless, with one degree of freedom, this approximation remains acceptable, and we verify that our histograms of likelihood values do not significantly depart from Gaussian behavior.

\section{Molecular mapping applied to MIRI MRS data}
\label{sec:mm_application}
\subsection{Correlation maps with the atmospheric template spectrum}
\begin{figure*}[h]
    \centering
    \includegraphics[width=18cm]{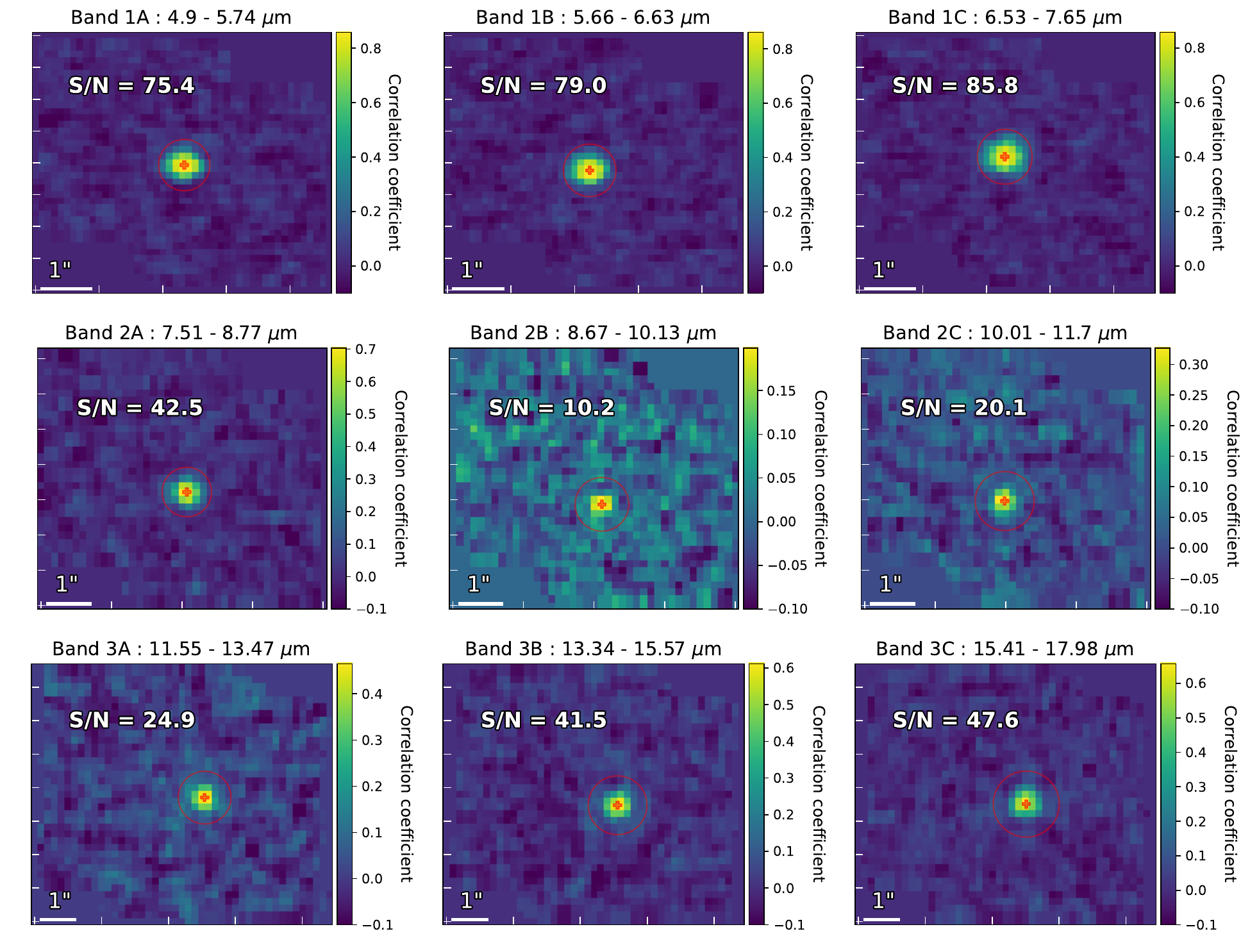}
    \caption{Correlation maps in each MRS band.
    The red cross indicated the position of the centroid of the PSF, previously identified from the median cubes.}
    \label{fig:corr_maps}
\end{figure*}

We start by calculating the correlation between the VHS\,1256\,b MRS data cubes and the atmospheric parameters that best fit the observations, using the method described in Sect.~\ref{sec:cross_correlation}.
The \texttt{Exo-REM} model used corresponds to the best-fit parameters $T = 1100$\,K and $\log g = 4.5$ \citep[Fig.~\ref{fig:ExoREM_molec_templates},][]{miles_jwst_2023}.
Metallicity and the $\mathrm{C/O}$ ratio are set to solar values at first.
The resulting correlation maps with this model are shown in Fig.~\ref{fig:corr_maps}.
The planetary-mass companion VHS\,1256\,b is detected with high $S/N$ in all bands.
The detection is lower in bands 2B, 2C and 3A compared to the other bands. 
However, before correlation and filtering, VHS\,1256\,b is visible in all channels with a decreasing flux with increasing wavelengths.
Bands 2B and 3A ($\sim$ 8.7-13.5 $\mu$m) correspond to the wavelengths of the silicate band \citep{miles_jwst_2023} and
silicate clouds attenuate molecular spectral signatures. 
The molecular mapping method appears to be less efficient for cloudy objects \citep[as already assumed in][]{cugno_molecular_2021}.
This is consistent with the findings that silicate absorption in young objects extends up to $\sim$13\,$\mu$m, whereas in older, cloudier objects it is confined to $\sim$11\,$\mu$m \citep{suarez_ultracool_2023}. 

Alternatively, this indicates that the models provide a poorer match to the data at these wavelengths, leading to lower correlation values.
Maximum correlation values in the spectral bands from 4.9 to 7.65\,$\mu$m (channel 1 bands) reach approximately 0.9, indicating an excellent agreement between the model used for correlation and the data.
In contrast, band 2B (8.7 - 10.1\,$\mu$m) shows a maximum correlation at the planet’s position of only $\sim$0.2, clearly reflecting a poor match.

\subsection{Detection of molecules at mid-infrared wavelengths}
\label{sec:mm_app_molecules}
\begin{figure*}[h]
    \centering
    \includegraphics[width=18cm]{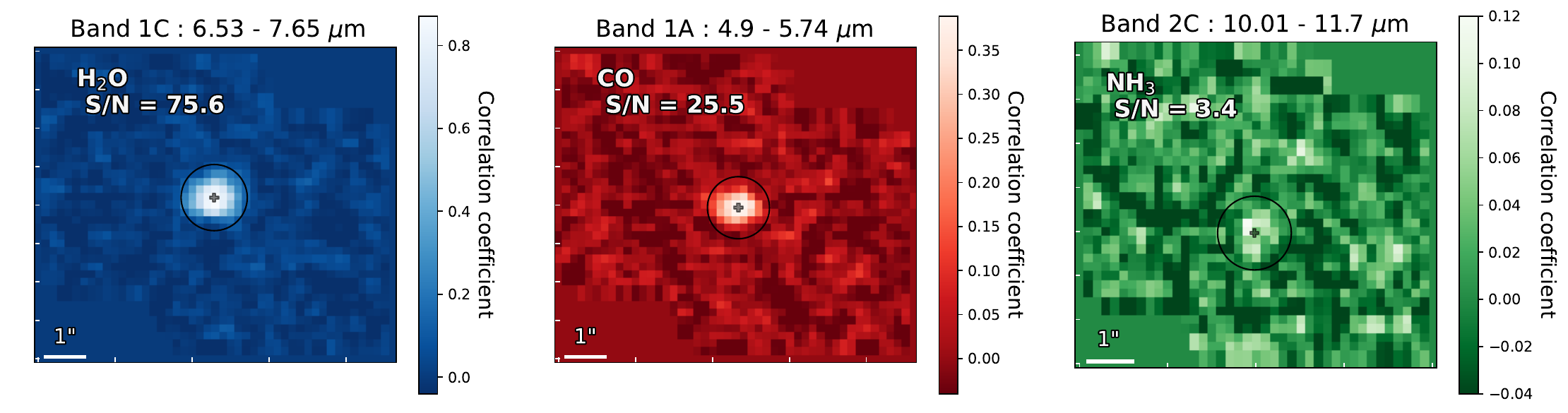} 
    \caption{Correlation maps with molecular templates presented in the band where the detection is the highest. The position of VHS\,1256\,b measured in the cubes is indicated by the black cross.}
    \label{fig:corr_maps_molec}
\end{figure*}

The data are then correlated with molecular templates, corresponding to a single atmospheric species, individually instead of the full atmospheric model (see Fig.~\ref{fig:ExoREM_molec_templates}).
The detected molecules are presented in Fig.~\ref{fig:corr_maps_molec} for the bands providing the clearest detections (see Fig.~\ref{fig:spectrum_bands_templates} for a version highlighting the spectral regions of such bands).
The corresponding histograms of the pixel value distributions are presented in Fig.~\ref{fig:histogram_cc_maps}.

The molecules H$_2$O and CO are robustly detected at the wavelengths where their spectral features are most prominent.
H$_2$O is detected with high confidence ($S/N$  = 24 --  76) in all bands, but lower $S/N$ in bands 2B and 2C ($S/N$ = 8 -- 17), as observed in the correlation maps with the full atmospheric model.
CO is detected only in band 1A ($S/N$ = 25.5).
Finally, NH$_3$ shows a tentative detection with a lower $S/N$ = 3.8 -- 3.1 in band 2B and 2C.
No other molecules are detected among those included in \texttt{Exo-REM}.

We note that CH$_4$ could be expected, since this molecule has previously been detected in VHS\,1256\,b's atmosphere in low abundance at shorter wavelengths \citep{miles_methane_2018}.
This molecule exhibits strong mid-IR spectral features ($\sim$ from 7 to 9 $\mu$m), 
and it is typically observed in L/T transition objects, particularly early-T types; but VHS1256\,b may be too warm to show CH$_4$ features in the mid-IR.
We note that we still measure a $S/N$ = 6 in band 2A using a CH$_4$ model spectrum. 
However, this apparent detection is biased by pixel-to-pixel artifacts in the MRS cube, as discussed in Sect.~\ref{sec:limitations}.
%
%
The correlation maps obtained with a featureless spectrum reveal a spurious signal with $S/N = 3.1$ at the planet’s position in band 2A, indicating a bias in this spectral band that complicates the confirmation of genuine detections. 
%
%
The case of NH$_3$ is also noteworthy, as it represents a marginal detection (Fig.~\ref{fig:corr_maps_molec}, right). 
For this molecule, the correlation with a featureless spectrum does not produce any spurious signal (Fig.~\ref{fig:artifacts}).

In order to test the reliability of all detections, we can subtract the correlation map obtained with a featureless spectrum from the ones computed with the molecular model we aim to study. 
For all VHS\,1256\,b correlation maps, the corrected maps show no $S/N$ values above 3 in spectral bands where no detection is expected. 
Applying the correction to the NH$_3$ maps does not change the correlation values, indicating that noise does not significantly affect the measured signal.
In contrast, the $S/N$ of the CH$_4$ correlation map decreases to $S/N = 3$ after correction, suggesting that the apparent CH$_4$ $S/N$ was increased by correlation with noise.
To further validate the NH$_3$ detection, we performed MIRISim simulations \citep[exactly as in][]{malin_simulated_2023} using model spectra to estimate the expected signals and identify which molecules could be detectable for given elemental abundances and exposure times.
The simulations confirmed that NH$_3$ is detectable, albeit weakly, even in the presence of silicate clouds in the models. 

Finally, the photocenter of the planet’s PSF does not always coincide with the spaxel with the highest correlation value, especially at lower $S/N$ detections.
The peak correlation does not necessarily correspond to the determined position of VHS\,1256\,b, especially in cases of lower $S/N$, as seen in the correlation map in band 2B and 2C in Fig.~\ref{fig:corr_maps}, and band 2C with NH$_3$ in Fig.~\ref{fig:corr_maps_molec}.
The crosses (red in Fig.~\ref{fig:corr_maps} and black in Fig.~\ref{fig:corr_maps_molec}) represent the measured centroids on the cubes, whereas the correlation maximum does not align with the same position in instances of low $S/N$.
The highest correlation value in the field of view does not precisely match the position of maximum flux in the cubes by approximately 1 pixel.
Given the limited spatial resolution \citep[0.196$''$ to 0.245$''$ per pixel][]{wells_mid-infrared_2015}, this cannot correspond to any physical effect; we instead suspect that it arises from the noise distribution (Sect.~\ref{sec:limitations}).

\section{1D cross-correlation using the extracted VHS\,1256\,b spectrum}
\label{sec:ccf_1d}

\subsection{Cross-correlation with molecular template}
\begin{figure*}[h]
    \centering
    \includegraphics[width=18cm]{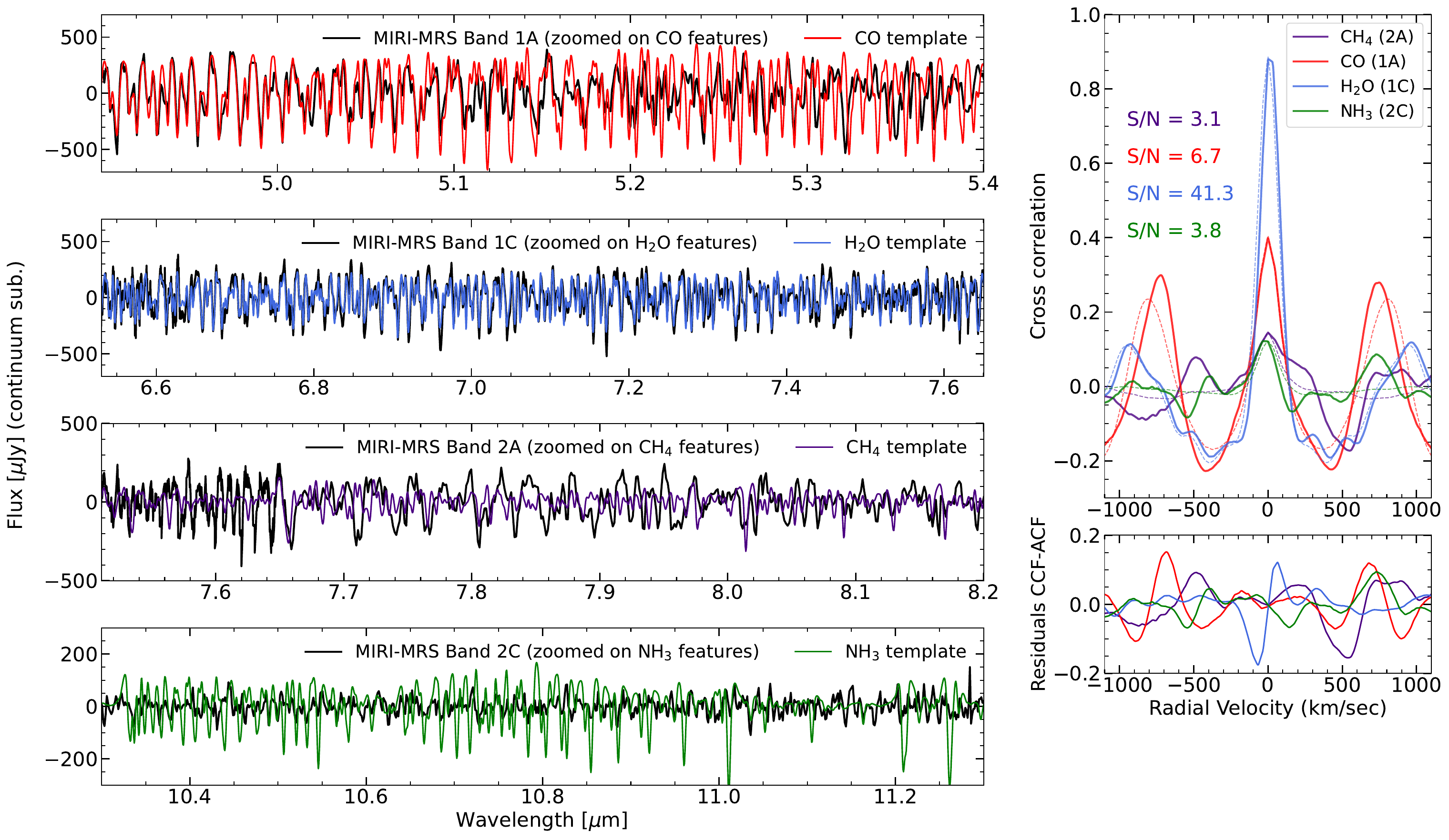}
    \caption{
    Left: Continuum-subtracted spectrum of VHS 1256\,b (black), shown with zooms on wavelength ranges corresponding to the features of the targeted molecules.
    The corresponding molecular templates are overplotted for comparison.
    Right:
    Cross-correlation functions with different molecular templates: H$_2$O, CO, NH$_3$, and CH$_4$, presented in the bands for which the $S/N$ is the highest. The autocorrelation functions (ACF) are shown with dashed lines.}
    \label{fig:ccf_molec}
\end{figure*}

We compute the CCF between the extracted spectrum in each band and the synthetic model spectra for each molecule.
Figure~\ref{fig:ccf_molec} shows the observed spectrum and the corresponding molecular templates in the associated spectral bands (see Fig.~\ref{fig:spectrum_bands_templates} for a zoomed-out version).
The CCF of the expected molecules are displayed on the right, each within the spectral band where the $S/N$ is the highest.
The $S/N$ values correspond to the method described in Sect.~\ref{sec:cross_correlation}. Results across all spectral bands are presented in Fig.~\ref{fig:ccf_molec_all_bands}.%
We verified that the measured correlations are not caused by inter-template cross-correlation among the molecular models. 
To do so, we computed the cross-correlation of the models with one another to confirm that they do not induce spurious signals, and we ensured that the detected molecular signal is consistent with that expected from atmospheric models of the targeted species.
The correlation with the full atmospheric model and with only H$_2$O signatures gives similar results in all bands, in agreement with the molecular mapping analysis.
The CCF with CO has a peak in the band 1A ($S/N$ = 6.7), with strong autocorrelation secondary peaks. 
The CCF with NH$_3$ have a peak with a $S/N$ from 3.0 to 3.6 in the band 2A and 2C, which is consistent with the results in molecular mapping.
Moreover, the CCF peak for CH$_4$ reaches $S/N = 3.1$ in Band 2A, providing a tentative detection of CH$_4$, which was not clearly identified with molecular mapping due to noise correlations that rendered the result unreliable.
%
We correlated the data with all molecular species included in \texttt{Exo-REM}, and note that no additional species were detected. 

Furthermore, we combine the spectra across the entire wavelength range (4.9--18 $\mu$m) to measure the CCF while varying the spectral window.
We find that computing the CCF using different spectral windows yields comparable results and does not improve the $S/N$ relative to analyzing the spectral bands individually (see Fig.\ref{fig:CCFs_entire_wave_range}).
%
Expanding the wavelength range does not enhance CH$_4$ detection compared to the observations in band 2A. 
The result obtained from CCF aligns consistently with the molecular mapping results, the latter providing greater $S/N$ in general.
The noise measured spatially in the correlation maps is lower than the noise measured in the radial velocity range from the CCF.
We intuitively suggest that this effect arises because molecular mapping builds noise estimates from multiple realizations of the spectrograph noise across different spatial locations, whereas 1D cross-correlation is restricted to a single ensemble of spaxels: the spectrum is unique, and the noise can only be assessed along the velocity axis.
The spatial extent of the correlation footprint, which can exceed the PSF size in cases of high S/N detection, allows the signal from multiple spaxels to be combined, which increases the resulting S/N.

\subsection{Cross-correlation with a self-consistent atmospheric grid}
\label{sec:exorem_grid_ccf}

\begin{figure}[h]
    \centering
    \includegraphics[width=8.6cm]{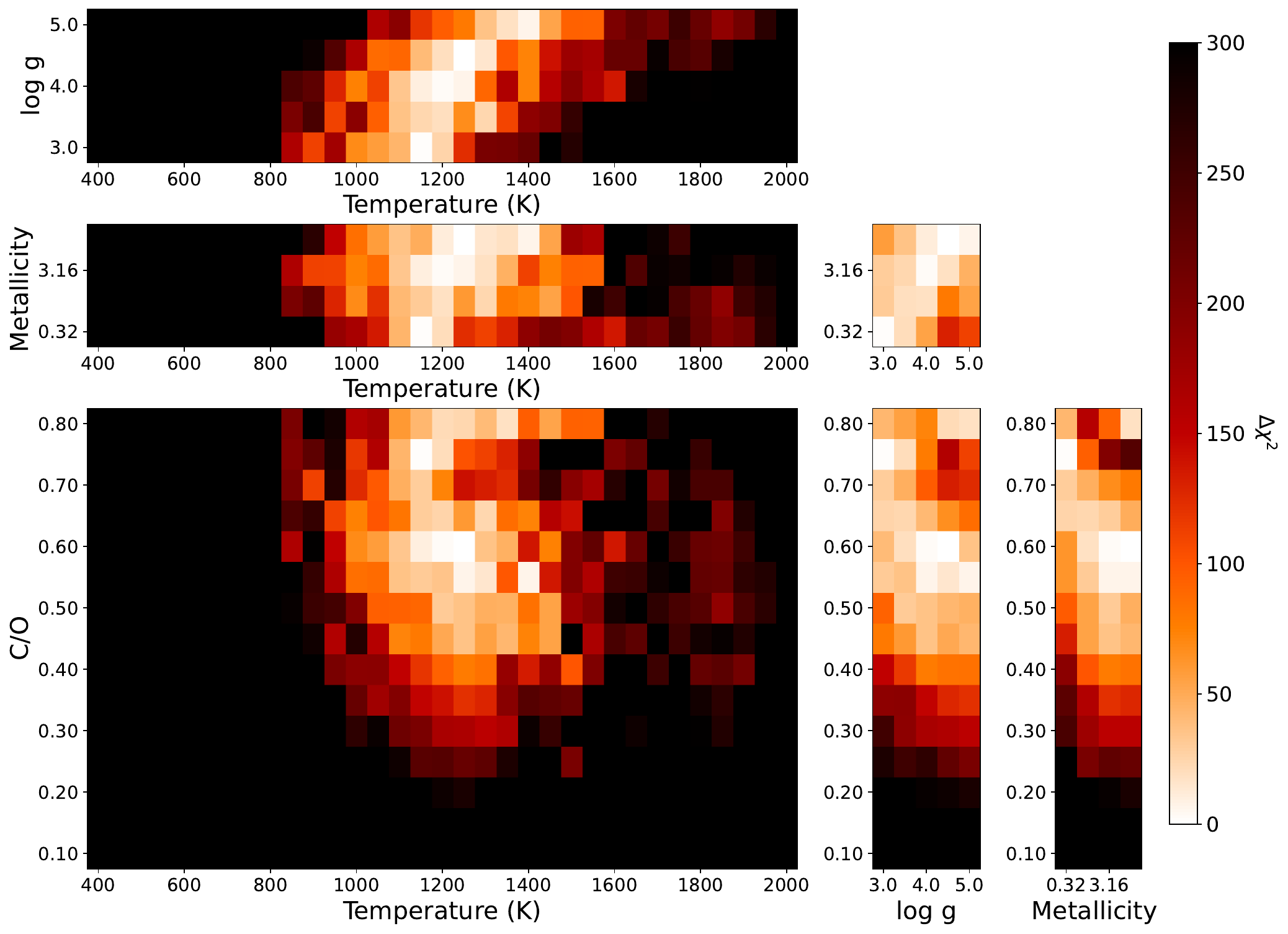}
    \includegraphics[width=8.6cm]{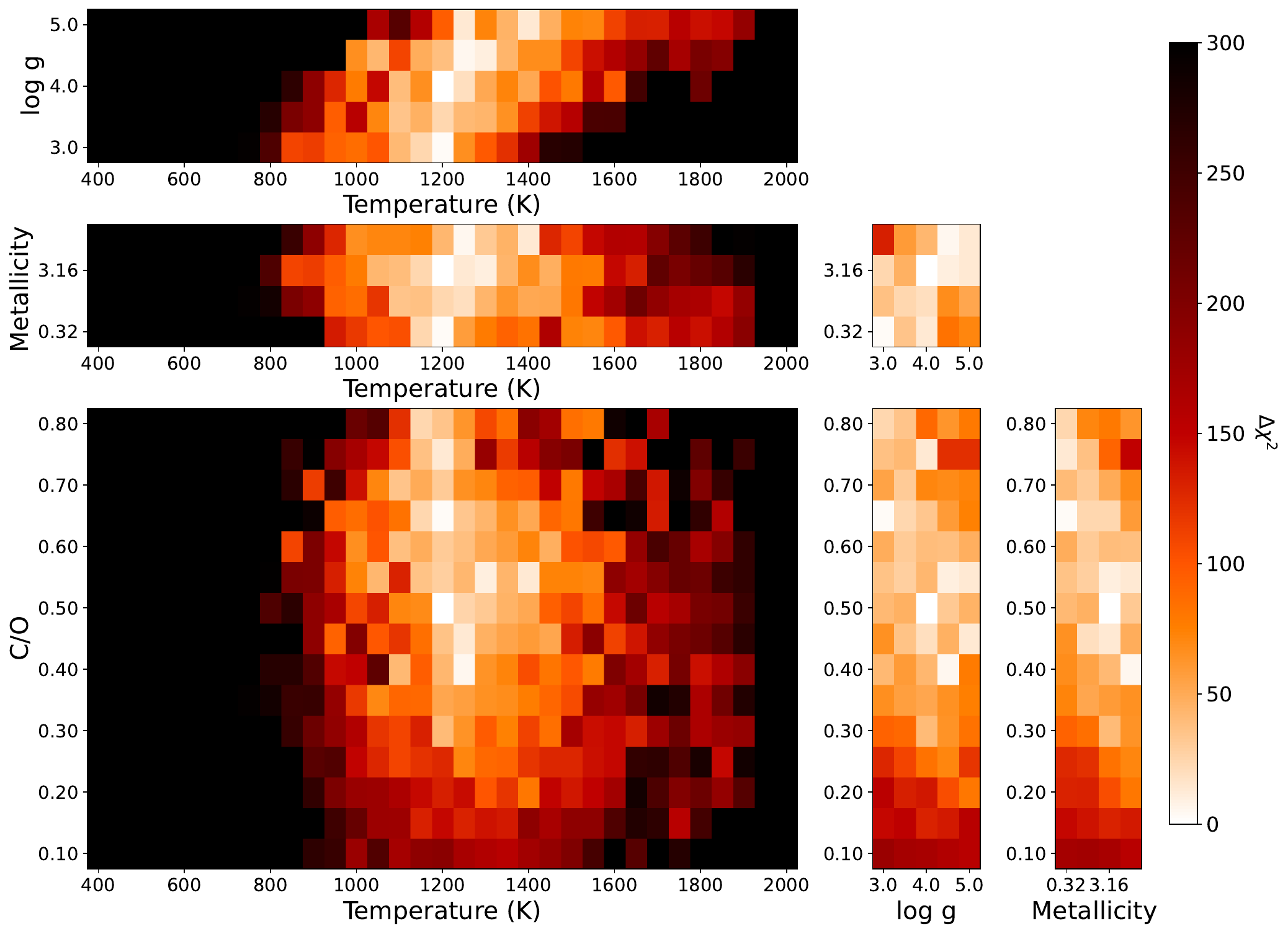}
    \caption{
    Grids of the correlation values with the \texttt{Exo-REM} models, converted into the $\Delta\chi2$. 
    Metallicity is expressed relative to the solar value $Z/Z_{\odot}$.
    Color scales are the same in each subplot.
    Top 3 panel : Band 1A (4.90--5.74 $\mu$m). 
    Bottom 3 panel: All MRS wavelength range (4.90--17.98 $\mu$m).}
    \label{fig:correlation_grid}
\end{figure}

Correlation methods have proven effective in detecting molecular species but are limited in accurately constraining the atmospheric parameters of planetary objects \citep{malin_simulated_2023}.
Once the low-frequency spectral components were removed from both the data and model spectra using a Gaussian filter to subtract the pseudo-continuum, the observed spectrum is correlated with each model in the grid.
We compute the correlation with every model from the \texttt{Exo-REM} atmospheric grid while varying four bulk properties: temperature, surface gravity, metallicity, and the $\mathrm{C/O}$ ratio of the model spectrum.
A substantial range of models yields high correlation values ($>0.8$), resulting in broad peaks in correlation space. 
We converted the correlations into $\chi^2$ values using Eq.~\ref{eq:cc-to-likelihood} and compared the correlations of different models.
%
Fig.~\ref{fig:correlation_grid} shows the $\Delta\chi^2$ between the best-fit models and all other models.
Each sub-panel displays the minimum $\chi^2$ for a given pair of parameters, while the remaining two parameters are varied. 
In practice, for each point in the grid corresponding to a specific parameter pair, we display the minimum $\chi^2$ (the maximum correlation coefficient) obtained when varying the other two parameters.

This measurement is first applied independently to each spectral band and then to the full spectrum (Fig.~\ref{fig:correlation_grid}, top 3 panels).  
Using the entire MRS wavelength range yields significantly tighter posterior distributions for the temperature and more precise constraints on the surface gravity, demonstrating that broader spectral coverage improves the determination of bulk atmospheric parameters (Fig.~\ref{fig:correlation_grid}, bottom three panels). 
However, the uncertainties on metallicity and the C/O ratio remain similar, or are even slightly larger, when the entire MRS range is included.
The highest correlation in Band~1A is obtained for the model with 
$T = 1250~\mathrm{K}$, $\log g = 4.5$, Z/Z$_\sun$ = 10, and $\mathrm{C/O} = 0.6$, with the $1\sigma$ uncertainties smaller than the grid resolution
(at $3\sigma$, the corresponding uncertainties are: 
$T = 1250 \pm 250~\mathrm{K}$, 
$\log g = 4.5 \pm 2.0$, 
Z/Z$_\sun$ = $10.0 \pm 9.68$, 
$\mathrm{C/O} = 0.6 \pm 0.2.$).
Using the entire wavelength range, the best-fit model is
$T = 1200~\mathrm{K}$, $\log g = 4.0$, Z/Z$_\sun$ = 3.16, and $\mathrm{C/O} = 0.5$.  
The $1\sigma$ uncertainties are smaller than the grid resolution (while at $3\sigma$ they are:
$T = 1200 \pm 50~\mathrm{K}$,
$\log g = 4.0 \pm 1.5$,
Z/Z$_\sun$ = $3.16 \pm 9.68$,
$\mathrm{C/O} = 0.5 \pm 0.25$).
Fig.~\ref{fig:Comparison_correlation_formosa} presents the comparison for each band, alongside the results of \citet{petrus_jwst_2024}.
The best-fit atmospheric parameters, particularly $\mathrm{C/O}$ and metallicity, depend on the spectral range considered.
The model yielding the highest correlation varies between bands, consistent with the findings of \citet{petrus_jwst_2024}.

Yet, as shown with the simulations \citep{malin_simulated_2023}, it is difficult to use a single band to fully resolve the degeneracies between some bulk parameters, surface gravity and metallicity, for instance.
On the contrary, the abundance of molecules, such as H$_2$O, CO, and CH$_4$, which define the $\mathrm{C/O}$ ratio, depends on the temperature of the planet.
Even though the pseudo-continuum of the planetary spectrum is lost in the filtering, those molecules still have a net effect in the high-frequency spectrum of the planet, explaining the relatively good match obtained for the temperature and the $\mathrm{C/O}$ ratio.
The other spectral bands produce similar results, but accuracy declines with increasing wavelength due to the reduced number of spectral features at longer wavelengths.

There are two hypotheses to explain the apparent degeneracies between some parameters, directly inherent to the cross-correlation method itself.
First, the noise floor could be too high, so the full line depth is not observable. 
While this is definitely the case when the exoplanet signal is buried in speckle noise, it cannot be the main explanation for VHS\,1256\,b, given the high statistical significance of the detection. 
Second, the relative depth of the lines could be affected by the high-pass filtering stage, thus limiting our ability to break the degeneracies between bulk properties.
We also note that model uncertainties could affect the results, including inaccuracies in line lists or templates generated from imperfect temperature–pressure profiles.

We retrieved the maximum correlation values from the 2D correlation maps and from the peak of the 1D CCF.
The results are similar whether using the correlation map at the planet’s spaxel or the 1D CCF peak, as expected; in a high $S/N$ regime such as this, no significant differences are anticipated between 1D and 2D cross-correlation.

The uncertainty in the estimated properties parametrized in the \texttt{Exo-REM} model presented here is smaller than the scatter across MRS channels presented in \cite{petrus_jwst_2024}.
Using the high-pass filtered version of the spectrum removes most degeneracies associated with thermal emission from clouds.
Clouds mostly modulate the continuum, which might be one of the limitations of the work presented in \cite{petrus_jwst_2024} as the \texttt{Exo-REM} model does not fully capture the cloud properties in L--T transition objects (Whiteford et al. subm).

\section{Assessing the presence of low-abundance molecules and isotopologues}

\label{sec:measuring_abundance}
\subsection{Measuring NH$_3$ abundance using cross-correlation}
\label{sec:nh3_abundance}
The standard approach to determining molecular abundances relies on atmospheric retrieval techniques. 
For VHS\,1256\,b, Whiteford et al. (subm) constrained the NH$_3$ volume mixing ratio VMR using NIRSpec and MIRI data, finding $[\rm NH_3] = -5.86^{+0.27}_{-0.23}$.
We investigate an alternative approach for estimating molecular abundances based solely on correlation values and a grid of forward models.
We construct a dedicated grid of 22 models in which only the [NH$_3$] is varied, with VMRs ranging from -4 to -9.3.
This grid assumes $[\rm H_2O] = -3.23$, and we note that alternative assumptions for the water VMR would impact the results.
We adopted a super-solar metallicity for the model grid, with $Z/Z_{\odot} = 3.16$, as indicated by the best-fit model spectrum.
To evaluate the sensitivity of our results to this assumption, we also computed a comparison grid assuming solar metallicity $Z/Z_{\odot} = 1$.
An additional model spectrum with no NH$_3$ is included for comparison.
These models are shown in Fig.\ref{fig:nh3_spectra}.

\begin{figure}[t]
    \centering
    \includegraphics[width=9cm]{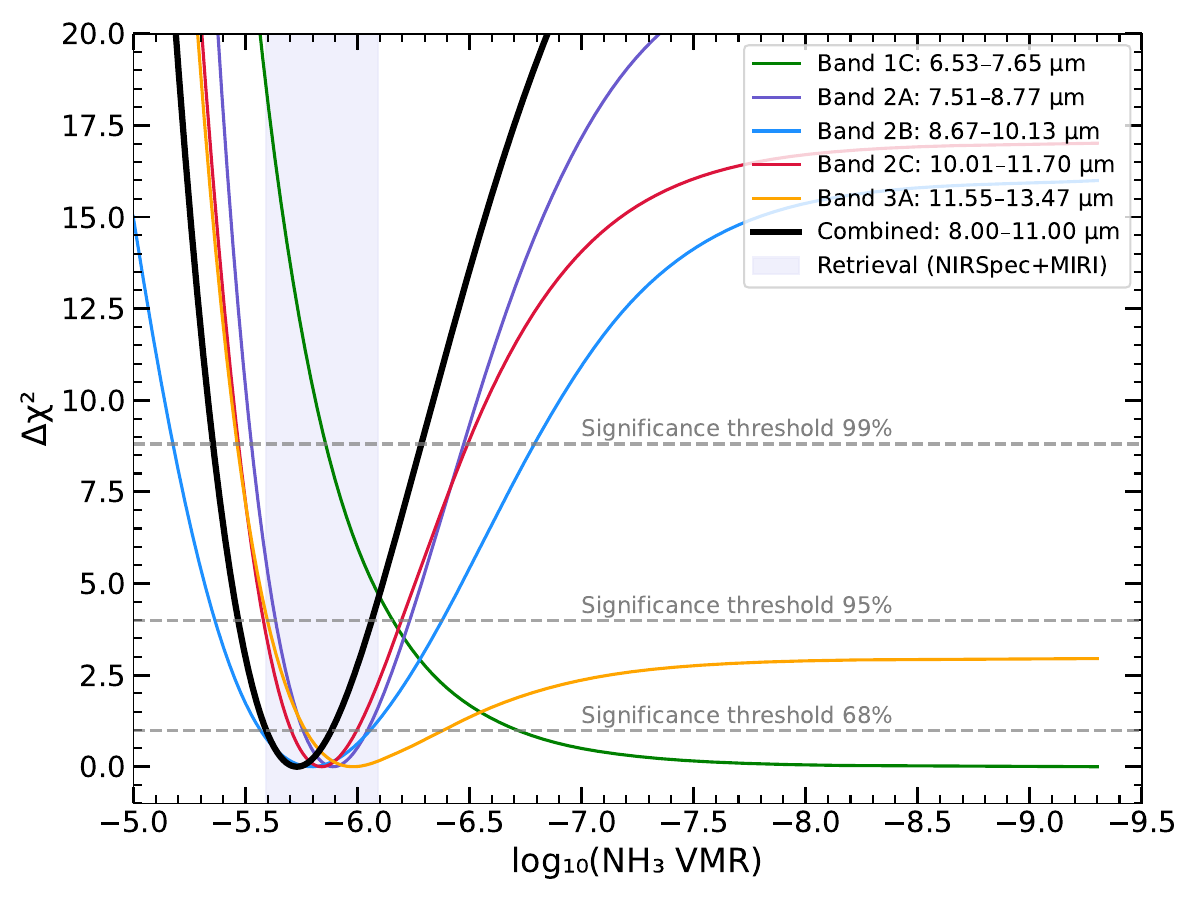}
    \caption{$\Delta\chi^2$ values computed for models with varying NH$_3$ abundances across MIRI bands 1C to 3A. 
    The black curve shows the results computed from 8--11 $\mu$m.
    For comparison, the NH$_3$ abundances derived from retrieval analysis is shown in background in purple (Whiteford et al. subm).}
    \label{fig:corr_vmr_nh3}
\end{figure}

We measure the correlation maps between the data cubes and each model, and we retrieve the highest correlation value at the planet's position in each band.
As a comparison, we also estimate the CCF using the VHS\,1256\,b spectrum.
We use both the CCF peak and the correlation map peak to qualitatively assess the statistical significance of correlation value variations as a function of model.
In band 1A, the correlation values vary by less than $\sim$0.1\% between the model with the highest correlation and the model without any NH$_3$.
However, in bands where NH$_3$ is expected, the correlation shows larger variations, up to 2\%.
Fig.~\ref{fig:corr_vmr_nh3} shows the $\Delta \chi^2$ values for each model.
These $\Delta \chi^2$ values are then interpolated over a fine grid covering a narrower range of NH$_3$ VMRs.
We present the results for each spectral band where NH$_3$ features could be visible and where the $\chi^2$ curves indicate a detection and converge toward a well-defined [NH$_3$].
Band 1C, which exhibits the fewest NH$_3$ features, is shown for comparison but does not provide meaningful constraints. 
Bands 2A to 3A allow reliable measurements of [NH$_3$], 
with abundances from individual bands consistent within 1-$\sigma$.
Combining the planet’s spectral information from 8–11 $\mu$m, where NH$_3$ features dominate, yields the strongest constraints.
We measure $[\rm NH_3] = -5.73^{+0.15}_{-0.14}$ (1-$\sigma$ uncertainties), consistent with values obtained from the retrievals.
Using the grid assuming solar metalicity, we measure $[\mathrm{NH_3}] = -6.44^{+0.14}_{-0.16}$.
We also note that we observe identical results when retrieving correlation values from correlation maps (i.e., on a single spaxel) and measuring the CCF with the extracted spectra (i.e., the observed spectrum of the object).

\subsection{Measuring the ratio $^{12}$CO/ $^{13}$CO}
\label{sec:isotopicCO_ratios}
\begin{figure}[t]
    \centering
    \includegraphics[width=9cm]{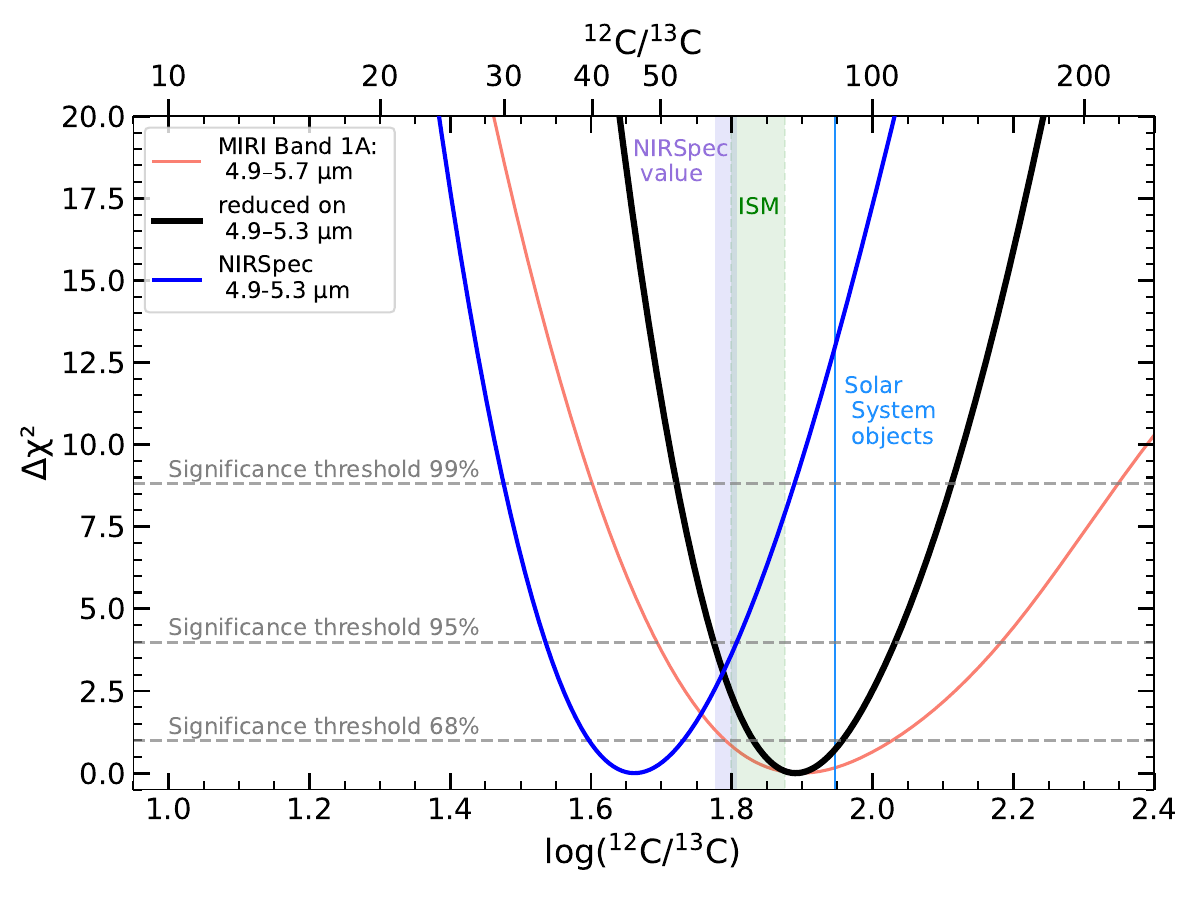}
   \caption{$\Delta\chi^2$ values computed for models with varying isotopic ratios of CO across different wavelength ranges. 
   For comparison, the isotopic ratios inferred from retrieval analyses are shown in light purple \citep{gandhi_jwst_2023}.
   The $^{12}\mathrm{C}/^{13}\mathrm{C}$ ratio from the local interstellar medium (ISM) is plotted in green \citep{wilson_isotopes_1999},
   and the corresponding ratio for solar system objects is plotted in blue \citep{woods_carbon_2009}.}
\label{fig:corr_ratiosCO}
\end{figure}

The $^{13}$C isotopologue has been detected at near-IR wavelengths (4.1--5.3\,$\mu$m) using NIRSpec G395H/F290LP, and a $^{12}$C/$^{13}$C ratio of 62$^{+2}_{-2}$ was previously measured \citep{gandhi_jwst_2023}.
We now investigate the feasibility of measuring the $^{12}$C/$^{13}$C isotopic ratio at MIRI wavelengths using a grid of \texttt{Exo-REM} models, following the same methodology as that adopted for the [NH$_3$] (Sect.~\ref{sec:nh3_abundance}).
We compute \texttt{Exo-REM} model spectra as described in Sect.~\ref{sec:pres_exorem}, with $\log_{10}(^{12}\mathrm{C}/^{13}\mathrm{C})$ varying from 1.0 to 2.6.
In all models, the total CO abundance was held fixed, such that the combined contribution of $^{12}$CO and $^{13}$CO remains constant.
For example, a value of $\log_{10}(^{12}\mathrm{C}/^{13}\mathrm{C}) = 1$ corresponds to a mixture of approximately 90\% $^{12}$CO and 10\% $^{13}$CO.
Our analysis focuses on the shortest MIRI wavelengths, which encompass the relevant CO spectral features.
The observed data are correlated with each model in the grid while varying the isotopic CO ratio.
The results are shown in Fig.~\ref{fig:corr_ratiosCO} for three cases.
Using the full MIRI Band~1A from 4.9--5.7 $\mu$m yields
$^{12}\mathrm{C}/^{13}\mathrm{C} = 79^{+28}_{-17}$.
When restricting the analysis to the 4.9--5.3 $\mu$m range, where we observe the most prominent CO features, we obtain 
$^{12}\mathrm{C}/^{13}\mathrm{C} = 77.8^{+13}_{-10}$.
Focusing on the strongest lines results in a consistent value but smaller uncertainties as compared to using the entire MIRI MRS Band 1A.
Overall, MIRI appears to yield higher $^{12}$C/$^{13}$C ratios, and we note that the derived isotopic measurement depends on the wavelength range considered.

For comparison, we applied the same analysis to the NIRSpec data (4.1--5.3 $\mu$m) to enable a direct and consistent comparison with previous results and this yields 
$^{12}\mathrm{C}/^{13}\mathrm{C} = 57^{+9}_{-8}$ \citep[using directly the NIRSpec spectrum from][]{miles_jwst_2023}.
The values obtained from both methods — the full retrievals and the correlation analysis based on model grids varying a single parameter — are consistent within 1$\sigma$ uncertainties, although our method yields a slightly lower value here but within error bars
($^{12}\mathrm{C}/^{13}\mathrm{C} = 57^{+9}_{-8}$ vs. $^{12}\mathrm{C}/^{13}\mathrm{C} = 62 \pm 2$).
This agreement, further supported by the NIRSpec-based results, demonstrates that the continuum-subtracted spectrum can provide a reliable estimate of the isotopic ratio.
Nonetheless, 
potential biases may arise from the models, the data, or the method itself, all of which could affect the inferred $^{12}\mathrm{C}/^{13}\mathrm{C}$ ratio. 
Moreover, even if the method performs well in this case, its robustness may not extend to other objects or spectral types, as it can depend on factors such as the target brightness, the strength of the CO signal, and the contrast with the host star in the case of exoplanets.
We used both the correlation values from the correlation maps and the peaks of the CCF, and we find that both methods yield consistent results within the uncertainties.

\section{Discussion}
\label{sec:discussion}

\subsection{Limitations of the molecular mapping method}
\label{sec:limitations}
The analysis of early JWST MRS datasets for the direct spectroscopy of a planetary-mass companion highlights some challenges.
\paragraph{Astrometry.} 
As predicted with simulated data, the astrometry based on correlation maps is not perfectly reliable, and the correlation footprint does not have a similar shape and size as a PSF.
This effect has been shown with MIRI MRS simulated data and was explained by the impact of the stellar signal, resulting in an asymmetry in the correlation footprint \citep{malin_simulated_2023}.
However, in the case of VHS\,1256\,b, the observed shift cannot be of stellar nature, as it is negligible at this separation.
The shift seems to be due to the noise distribution and/or the presence of cosmic showers.
Therefore, noise can have an impact on detection in the case of low $S/N$, shifting the correlation maximum to an adjacent pixel.

\paragraph{Noise artifacts.}
Noise can induce spurious positive correlations, making detections more challenging at low $S/N$. 
In some cases, the $S/N$ may even be overestimated due to random correlations with noise at the planet’s position, potentially leading to false detections (Sect. \ref{sec:mm_app_molecules}).
Therefore, this effect must be carefully evaluated before confirming any signal.
One practical approach to assess its impact is to subtract the correlation map obtained with a featureless reference spectrum and verify that no residual noise correlations artificially enhance the measured $S/N$.
We note that the high-frequency noise has been identified to originate from the data reduction, in particular from straylight, although it has not yet been fully quantified \citep{martos_combining_2025}.
More in-depth methods will enable to better distinguish correlations arising from noise from those associated with the planetary signal.
This noise pattern may also arise from fringing and sampling-related oscillations, which can correlate with molecular template spectra and produce spurious signals.

\paragraph{Clouds.}
Correlation with exoplanet atmosphere grid models can be inefficient for studying cloudy or very young planets, whose atmospheres may be heavily obscured by clouds or dust, attenuating near-IR molecular features, as noted by \citet{cugno_molecular_2021} for the $\sim$5\,Myr protoplanet PDS\,70\,b.
Similar effects are expected at mid-IR wavelengths, where silicate cloud features can obscure or attenuate molecular spectral signatures.

\paragraph{Atmospheric properties.}
Assessing the atmospheric properties of planetary-mass companions using spectral cross-correlation can be challenging, as this technique relies primarily on the high-frequency spectral content \citep[e.g.][]{hoeijmakers_medium-resolution_2018, petrus_medium-resolution_2021}.
These previous studies, which relied solely on $K$-band spectra, have shown that the effective temperature (T$_{\mathrm{eff}}$) can be biased, 
as it has little influence on the high-frequency spectral features at these wavelengths. 
Our results show that the MIRI MRS spectral range provides improved sensitivity to T$_{\mathrm{eff}}$, 
thanks to molecules such as NH$_3$ (for T$_{\mathrm{eff}} \sim 800$--1200~K) and CH$_4$ at lower temperatures, which exhibit strong temperature-dependent features. 
The depth of the H$_2$O absorption feature at 6~$\mu$m is also correlated with the object’s temperature. 
Although some degeneracies between surface gravity and metallicity remain, 
these parameters are better constrained by the broad wavelength coverage offered by MIRI MRS, 
as demonstrated with simulated data \citep{malin_simulated_2023}.
The broad mid-IR spectral coverage thus provides improved constraints on T$_{\mathrm{eff}}$, even when using correlation-based analyses alone.
Molecular mapping is more effective for constraining relative abundances and, consequently, the $\mathrm{C/O}$ ratio, especially for L-type objects whose atmospheres are dominated by H$_2$O and CO \citep{petrus_medium-resolution_2021}. 
Our analysis suggests a solar to super-solar $\mathrm{C/O}$ ratio (C/O = 0.5 using the entire MIRI wavelength range), consistent with the findings of \citealt{petrus_x-shyne_2023, hoch_assessing_2023}.

\paragraph{Impact of model priors}
Cross-correlation relies on prior knowledge of the planet’s atmosphere. 
The relative insensitivity to variations in the template model is advantageous, allowing the planet to be detected even if the model does not perfectly reproduce its atmosphere.
The method remains effective as long as the same molecular features are present in both the models and the data.
Once the planet is detected, the data cube can be correlated against a full grid of atmospheric models to refine the atmospheric parameters.
\paragraph{Continuum estimation}
Another limitation arises from the precision with which the continuum is removed. 
If the continuum resolution is too low, not all the pseudo-continuum is subtracted, whereas if it is too high, useful information for the CCF may be removed.
Therefore, the choice of continuum resolution can significantly impact the resulting CCF.

\subsection{Detection of molecules and abundances measurement}
\paragraph{CH$_4$} 
The marginal detection of CH$_4$ at mid-IR wavelengths could be due to the strong H$_2$O absorption obscuring its signatures, along with silicate clouds that still diminish spectral features.

\paragraph{NH$_3$} 
Atmospheric \texttt{Exo-REM} model spectrum computed using the bulk atmospheric parameters of VHS\,1256\,b predict the presence of ammonia in its atmosphere, albeit at low abundance.
Consequently, the detection of NH$_3$ at only a low $S/N$ is fully consistent with these predictions.
The mid-IR wavelength range offers a significant advantage by targeting the prominent NH$_3$ absorption feature around 10~$\mu$m.
MIRI MRS can detect minor molecules at these wavelengths, such as NH$_3$ in L-type companions, particularly in this case without stellar contamination.
Since the NH$_3$ spectral signature is clearly distinct from stellar features, spectral cross-correlation should also be able to detect it in close-in companions even in the presence of host star contamination, as any correlation with stellar features is expected to be minimal.
Moreover, cross-correlation techniques can effectively probe the gas above the clouds, which primarily induce broad spectral modulations that are partly removed by high-pass filtering.

The method is particularly effective for trace species, allowing us to obtain meaningful constraints even when the molecule is detected at low $S/N$.
If NH$_3$ is detected with an $S/N$ of 3 in a band such as 2C, the corresponding uncertainty can be estimated as $\Delta \log[{\rm NH_3}] \approx 0.33$,
which is consistent with the values observed in Fig.~\ref{fig:corr_vmr_nh3}.
Nevertheless, the overall uncertainty remains tied to the metallicity and H$_2$O abundance. 
Since abundances generally scale with metallicity, the primary quantity being constrained is the $
[{\rm NH_3}]/[{\rm H_2O}]$ ratio. 
For a trace species like NH$_3$, partially masked by other molecular features, the signal is roughly proportional to its abundance (unlike for major species) motivating our specific test of its measurability.
The depth of the NH$_3$ lines increases with its abundance relative to H$_2$O and the continuum.

Ammonia becomes a more dominant species in atmospheres cooler than typical T types, and remains an abundant trace species up to relatively high temperatures, consistent with its clear detection in the mid-IR spectra of T-type brown dwarfs and exoplanets \citep{cushing_spitzer_2006, malin_first_2025, beiler_precise_2024}.
The onset of this feature is generally associated with the L/T transition \citep{cushing_spitzer_2006} and more specifically with spectral type T2.5 \citep{suarez_ultracool_2022}. 
Therefore, confirming its signature in VHS~1256~b indicates the presence of NH$_3$ in an object of earlier spectral type than previously anticipated.
 
\paragraph{CO}
We have shown that CO isotopologues are also detectable with MIRI MRS (particularly at the shortest MRS wavelengths). 
We obtained results that are consistent with the values measured with the NIRSpec instrument.
The CO isotopic ratio offers key insights into carbon enrichment during planet formation beyond the snowline.
The first measurement of the $^{12}$CO/$^{13}$CO ratio for the planetary companion YSES\,1\,b \citep[$^{12}$CO/$^{13}$CO = 31$^{+10}_{-17}$;][]{zhang_13co-rich_2021} suggested that this exoplanet have formed well beyond the CO snowline, accreting a significant fraction of its carbon from ices enriched in $^{13}$C. 
However, recent measurement using VLT/CRIRES$^{+}$ (R = 100,000) indicates $^{12}$CO/$^{13}$CO = 88$\pm$13 for this same object \citep{zhang_eso_2024}.
This discrepancy highlights the need for the mid-IR regime to measure CO isotopologues (4--6 $\mu$m), together with high spectral resolution.
For VHS\,1256\,b, we confirm an intermediate carbon isotopologue ratio, between the ISM value \citep[$^{12}\mathrm{C}/^{13}\mathrm{C} = 69 \pm 6$,][]{wilson_isotopes_1999} and that of solar system objects \citep[$^{12}\mathrm{C}/^{13}\mathrm{C} = 88.3 \pm 3.5$][]{woods_carbon_2009}.
Future NIRSpec G395H and MIRI MRS observations of planetary-mass companions should enable measurements of this ratio across a range of objects.
%
The regime probed by NIRSpec (3–5\,$\mu$m) encompasses the CO first overtone and fundamental rovibrational bands \citep{goorvitch_infrared_1994}, 
which trace deeper atmospheric layers than many mid‑IR features.
This makes it likely to yield a less biased measurement of the isotopic ratio.

\subsection{Ammonia as a probe for the mass}
\begin{figure}[t]
    \centering
    \includegraphics[width=8cm]{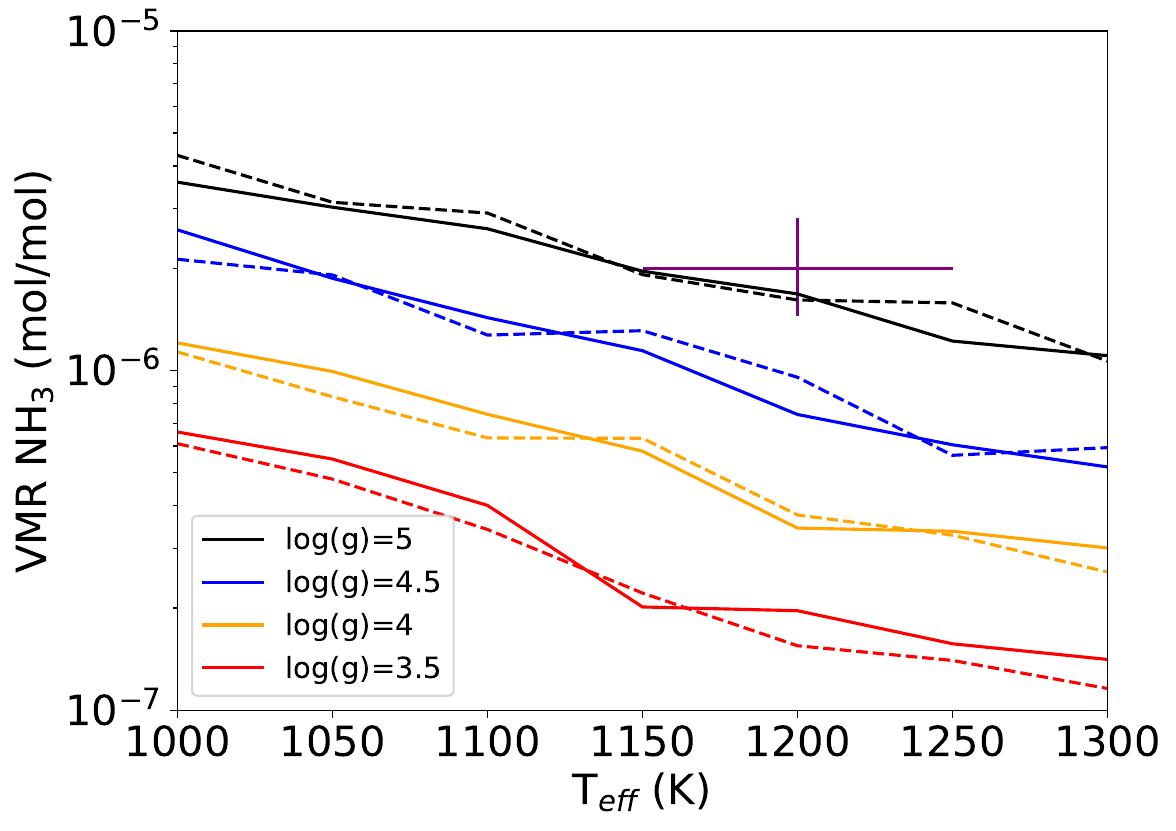}
    \caption{Volume mixing ratio (VMR) of NH$_3$ as a function of effective temperature for different surface gravity values, as measured with \texttt{Exo-REM}.
    Solid curves correspond to solar metallicity and dashed curves to 3 times solar metallicity. 
    The purple cross indicates measured values for VHS\,1256\,b (Sect. \ref{sec:nh3_abundance}).}
    \label{fig:vmr_nh3}
\end{figure}

Measuring the NH$_3$ abundance provides key insights into planetary atmospheres: while largely insensitive to vertical mixing and metallicity \citep{zahnle_methane_2014}, it is strongly temperature dependent and thus probes the deep atmosphere at 1 to 5 bar.
When the effective temperature and radius are well constrained, as for VHS\,1256\,b, NH$_3$ further constrains the surface gravity and hence the mass \citep{zahnle_methane_2014}, see Fig.~\ref{fig:vmr_nh3}.
The [NH$_3$] derived in our analysis indicate a high surface gravity of $\mathrm{log} g \geq 5.0$ for an effective temperature of T$_{\rm eff} = 1150$–$1250\,\mathrm{K}$.
For an assumed radius of 1.27\,R$_{\rm Jup}$ \citep{miles_jwst_2023}, this implies a mass exceeding 20\,M$_{\rm Jup}$.
We note that metallicity has only a minor influence on this estimate.
This result is consistent with the high mass end of the bimodal distribution predicted by evolutionary models \citep{miles_jwst_2023}. 
Incorporating the additional constraint from NH$_3$ favors the higher mass solution and places VHS\,1256\,b above the deuterium burning limit.
This measurement may be interpreted as favoring the higher-mass scenario; however, it remains strongly model-dependent.

\subsection{Future Perspectives}
We obtain consistent results using both the molecular mapping technique, without spectral extraction, and direct extraction of the companion spectrum from the data cube.
This validates spectral cross-correlation methods with MIRI MRS as a robust method for exoplanet atmospheric characterization, particularly when stellar contamination prevents spectral extraction, and demonstrates its effectiveness at mid-IR wavelengths with observations of a planetary-mass companion.

Observations of long-period exoplanets with MIRI-MRS are already scheduled and cover a wide range of planetary properties: 
Program GO 3647 \citep{patapis_gj504_2023} targets GJ\,504\,b, one of the coldest directly imaged companions; Program GO 4829 \citep{boccaletti_spectroscopic_2024} focuses on the HR\,8799 system, which hosts four giant planets with effective temperatures between 800 and 1200\,K; 
and Program GO 8714 observes the coldest directly imaged planet, Esp\,Ind\,b \citep{matthews_temperate_2024}, recently imaged with JWST and estimated to have a temperature of $\sim$275\,K \citep{sanghi_worlds_2026}.
Spectral cross-correlation will be particularly important in this higher-contrast imaging regime compared to VHS\,1256\,b, where extracting the continuum from the planetary signal is challenging.
For these upcoming observations, the molecular mapping method offers a direct way to detect planets, even when they are completely hidden in stellar light.
This method will enable the detection of molecules with high confidence without the need to extract a spectrum for the target, while also providing initial constraints on atmospheric parameters.
Notably, the spectroscopic properties and photometric colors of VHS\,1256\,b are similar to those of the four giant planets in HR\,8799 \citep{gauza_discovery_2015}, covering the L–T transition and allowing studies of effects of cloud.
The two other exoplanets, GJ\,504\,b and Eps\,Ind\,b, probe even cooler T-type atmopshere with rich mid-IR molecular features and detectable isotopologues.
This method is offering a powerful way to detect the full molecular content across a range of atmospheric temperatures.
%

The development of PSF subtraction methods \citep[e.g.,][]{worthen_miri_2024} effectively removes stellar contamination and enables the extraction of a low-resolution spectrum for $\beta$\,Pic\,b (at 0.5\,$''$ from the star).
The stability of space-based observations is promising for using such RDI-based techniques using either dedicated reference star observation or observation coming from former programs \citep[e.g.][]{cugno_mid-infrared_2024}.
Adapting the \texttt{breads} framework \footnote{\href{https://github.com/jruffio/breads}{https://github.com/jruffio/breads}} \citep{ruffio_jwst-tst_2024} to the MIRI-MRS instrument could enable more robust extraction of continuum-subtracted spectra for these planets.
The combination of PSF subtraction and cross-correlation methods is a promising avenue for characterizing exoplanetary atmospheres.

Upcoming IFUs with higher spectral resolution \citep[e.g., $R \sim 100,000$ for ELT-METIS,][]{brandl_metis_2021}
at similar wavelengths, as well as future instruments operating at different wavelengths
\citep[e.g., ELT-PCS,][]{kasper_pcs_2021}, will further expand these capabilities.
The next generation of ELT-class telescopes will enable the characterization of planets much closer to their host stars, where correlation-based methods are expected to provide a significant improvement in detection and atmospheric analysis.

In this work, we have employed a grid-fitting approach using cross-correlation, together with a conversion from correlation to likelihood.
Having demonstrated that this method performs reliably, it can now be extended to a more complete Bayesian forward-modeling framework based solely on correlation, and combining with photometry following, for example, \cite{brogi_retrieving_2019}. 
Similar approaches are beginning to be applied to more complex datasets, such as for $\beta$\,Pic\,b \citep{ravet_multimodal_2025}.
The conversion from correlation to likelihood used in this work appears consistent to first order, but requires further investigation, as the noise is not always perfectly Gaussian and the distribution of correlation values tends to be asymmetric, particularly at low $S/N$.

\section{Conclusion}
\label{sec:conclusion}
We conclude with a summary of the key findings of this work.
\begin{itemize}
\item 
We updated the data reduction using the latest calibration pipeline, which yields a clear improvement in $S/N$, particularly at the longest wavelengths.

\item 
We applied the molecular mapping method, developed on simulated data, and found that the predictions of the simulations are consistent with the observed MRS measurements \citep{patapis_direct_2022, malin_simulated_2023}.
This high S/N dataset for a planetary-mass companion with properties comparable to those of directly imaged exoplanets allows a direct comparison between molecular mapping on the data cube and correlation analyses using the extracted 1D spectrum. 
Both approaches yield consistent results: molecular mapping recovers the expected signals without requiring an explicit spectral extraction, and the analysis based on the extracted spectrum of VHS\,1256\,b leads to the same conclusions.

\item 
Correlation-based techniques provide a powerful framework for molecular detection, enabling both broad searches across many species and targeted tests for specific molecules. 
For VHS\,1256\,b, we robustly detect H$_2$O ($S/N \sim 76$) and CO ($S/N \sim 25$), and obtain tentative detections of NH$_3$ and CH$_4$ at lower significance ($S/N \sim 3$). While H$_2$O and CO are clearly recovered, less abundant species such as NH$_3$ are identified primarily through correlation, despite their spectral features being invisible in the observed spectra, consistent with independent free-chemistry retrieval analyses (Whiteford et al., subm.).
CH$_4$ is tentatively seen in the cross-correlation functions but is not clearly recovered in the correlation maps. 
The method also highlights the presence of clouds in the planet’s atmosphere.

\item 
We use cross-correlation with a self-consistent \texttt{Exo-REM} atmospheric model grid to constrain atmospheric parameters: effective temperature, surface gravity, metallicity, and carbon-to-oxygen ratio.
Using the full wavelength range, we derive $T = 1200 \pm 50~\mathrm{K}$, $\log g = 4.0 \pm 1.5$, Z/Z$_\odot = 3.16 \pm 9.68$, and $\mathrm{C/O} = 0.5 \pm 0.25$ (3$\sigma$ uncertainties).
These values are consistent with those derived from a range of various analysis methods.
The effective temperature and C/O ratio are tightly constrained, while the metallicity and surface gravity remain less well determined but consistent within the uncertainties at the 3$\sigma$ level.

\item 
Using cross-correlation with model grids, we measure a volume mixing ratio of $[\rm NH_3] = -5.73^{+0.15}_{-0.14}$, consistent with estimates from free-chemistry retrievals based on the combined MIRI and NIRSpec spectrum. 
We measure a $^{12}$C/$^{13}$C ratio of $77.8^{+13}_{-10}$ with MIRI, slightly lower but consistent with previous near-IR JWST/NIRSpec retrievals \citep{gandhi_jwst_2023}.
These results illustrate that correlations with atmospheric models can independently constrain the abundances of trace species and even isotopic ratios.

\item 
\texttt{Exo-REM} models indicate that the NH$_3$ volume mixing ratio is highly sensitive to the surface gravity of an atmosphere. 
For the NH$_3$ abundance measured for VHS\,1256\,b, we infer the corresponding surface gravity, which, combined with the retrieved temperature, implies a mass above the deuterium-burning limit.

\item 
This study shows that MRS mid-IR spectro-imaging is particularly promising for detecting and characterizing fainter, colder objects than VHS\,1256\,b, and for exoplanets in higher-contrast regimes (planet-star contrast of $\sim$ 10$^{-5}$ in the mid-IR; 
such as those targeted in upcoming observations of GJ\,504\,b, the HR\,8799 system, and the temperate companion Eps\,Ind\,b).
The approach naturally extends to even colder and intrinsically fainter companions that remain observable in the mid-IR, including recently identified companions such as TWA\,7\,b \citep{lagrange_evidence_2025} and 14\,Her\,c \citep{bardalez_gagliuffi_jwst_2025}, with exposure times at least 25 $\times$ longer than for VHS 1256 b to reach the faintness of such targets. (Exposure times range from over 8 hrs for TWA\,7\,b to 12 hrs for 14\,Her\,c to achieve a S/N above 10 per wavelength at their brightness peak.)
\end{itemize}

\begin{acknowledgements}
This work is based on observations made with the NASA/ESA/CSA James Webb Space Telescope. The data were obtained from the Mikulski Archive for Space Telescopes at the Space Telescope Science Institute, which is operated by the Association of Universities for Research in Astronomy, Inc., under NASA contract NAS 5-03127 for JWST. These observations are associated with program ERS 1386.
The specific observations analyzed can be accessed via \dataset[doi: 10.17909/17zp-8e91]{https://doi.org/10.17909/17zp-8e91}.
This work was granted access to the HPC resources of MesoPSL financed by the Region Ile de France and the project Equip@Meso (reference ANR-10-EQPX-29-01) of the programme Investissements d’Avenir supervised by the Agence Nationale pour la Recherche.
BJS acknowledges funding by the UK Science and Technology Facilities Council (STFC) grant nos. ST/V000594/1 and UKRI1196
A.Z. acknowledges support from ANID -- Millennium Science Initiative Program -- Center Code NCN2024\_001 and Fondecyt Regular grant number 1250249.
M.T. is supported by JSPS KAKENHI grant No.24H00242. J. M. V acknowledges funding from European Research Council Starting Grant Exo-PEA (Grant agreement No. [101164652]).
C.D. acknowledges financial support from the grant RYC2023-044903-I funded by the MCIU/AEI/10.13039/501100011033 and by the ESF+.

\end{acknowledgements}

\begin{contribution}
M. Mâlin led the data reduction, data analysis, scientific interpretation, and wrote the manuscript. 
A. Boccaletti, B. Charnay, L. Pueyo, and A. Bidot contributed to the data analysis and interpretation of the results.
B. Charnay computed the atmospheric models essential for the data analysis.
P. Patapis initiated some ideas that motivated this work.
S. Hinkley, S. Petrus, N. Whiteford and M. Perrin provided continuous feedback on the data interpretation.
All authors are members of the ERS team, reviewed the manuscript and provided comments.
\end{contribution}
\facilities{JWST (MIRI, NIRSpec)}
\software{\texttt{astropy} \citep{the_astropy_collaboration_astropy_2013, the_astropy_collaboration_astropy_2018,the_astropy_collaboration_astropy_2022}, \texttt{scipy} \citep{virtanen_scipy_2020},
        \texttt{numpy} \citep{harris_array_2020},
        \texttt{pandas} \citep{the_pandas_development_team_pandas-devpandas_2021, wes_mckinney_data_2010},
        \texttt{matplotlib} \citep{hunter_matplotlib_2007}}
\appendix

\section{\texttt{Exo-REM} molecular templates spectra}
\label{app:ExoREM_molec_templates}
\begin{figure}[h]
    \centering
    \includegraphics[width=9cm]{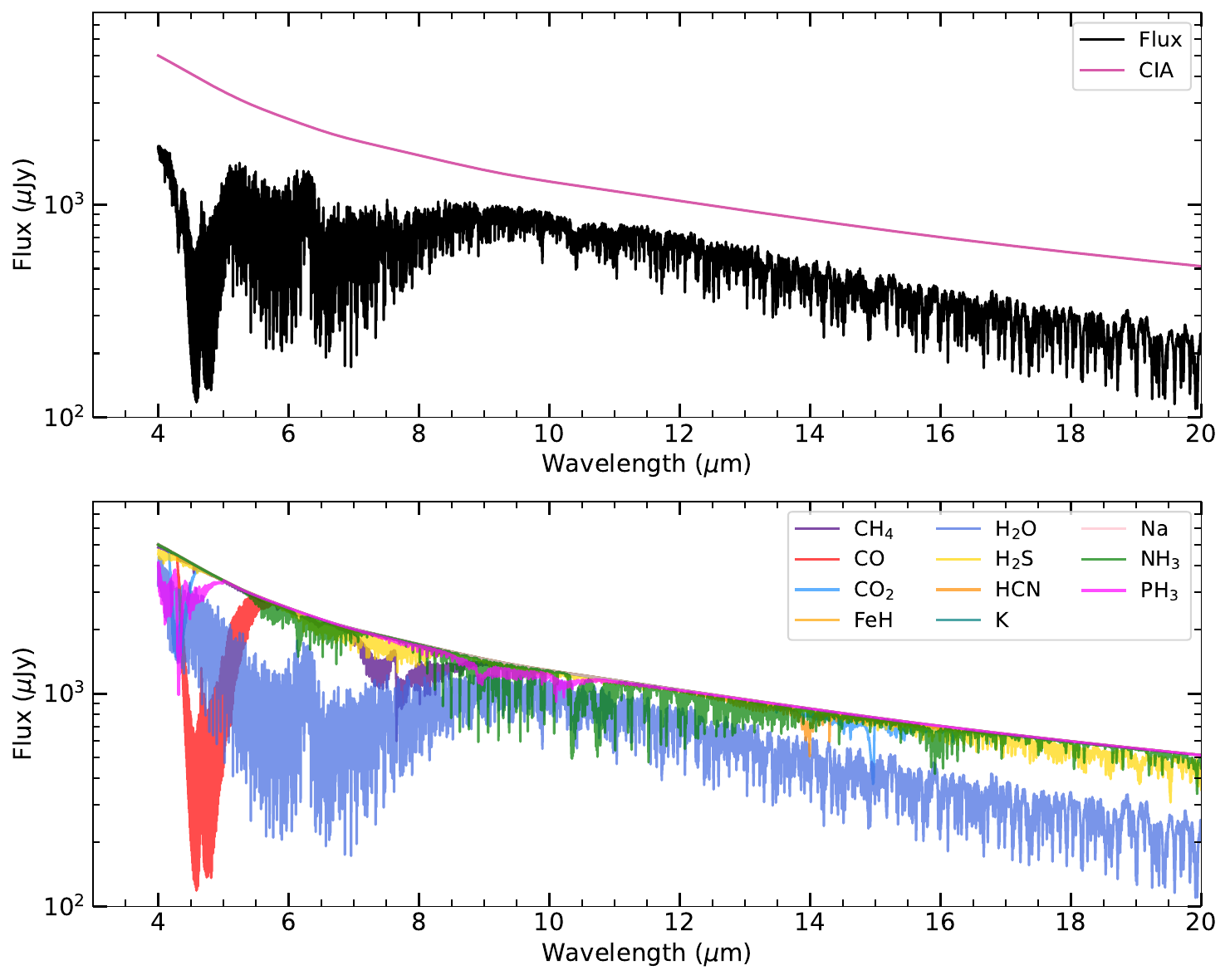}
    \caption{
    Top: \texttt{Exo-REM} molecular template spectrum used in the molecular mapping analysis. 
    The collision-induced absorption (CIA) continuum is also shown.
    Bottom: Each template corresponds to a different molecule whose detection was tested in the analysis.}
    \label{fig:ExoREM_molec_templates}
\end{figure}

This appendix presents the molecular template spectra used in our analysis. 
Figure~\ref{fig:ExoREM_molec_templates} shows all the molecules included in \texttt{Exo-REM} for which we investigated the presence of molecular signatures
We note that only CH$_4$, CO, CO$_2$, H$_2$O, H$_2$S, HCN, NH$_3$ and PH$_3$ exhibit visible spectral features at 1100\,K under the assumptions of these models.
Future refinements of the chemical network will further improve the accuracy of the self-consistent chemistry, for example by better capturing species such as PH$_3$, which is not detected at the level currently expected \citep{beiler_tale_2024}.

\begin{figure}[h]
    \centering
    \includegraphics[width=9cm]{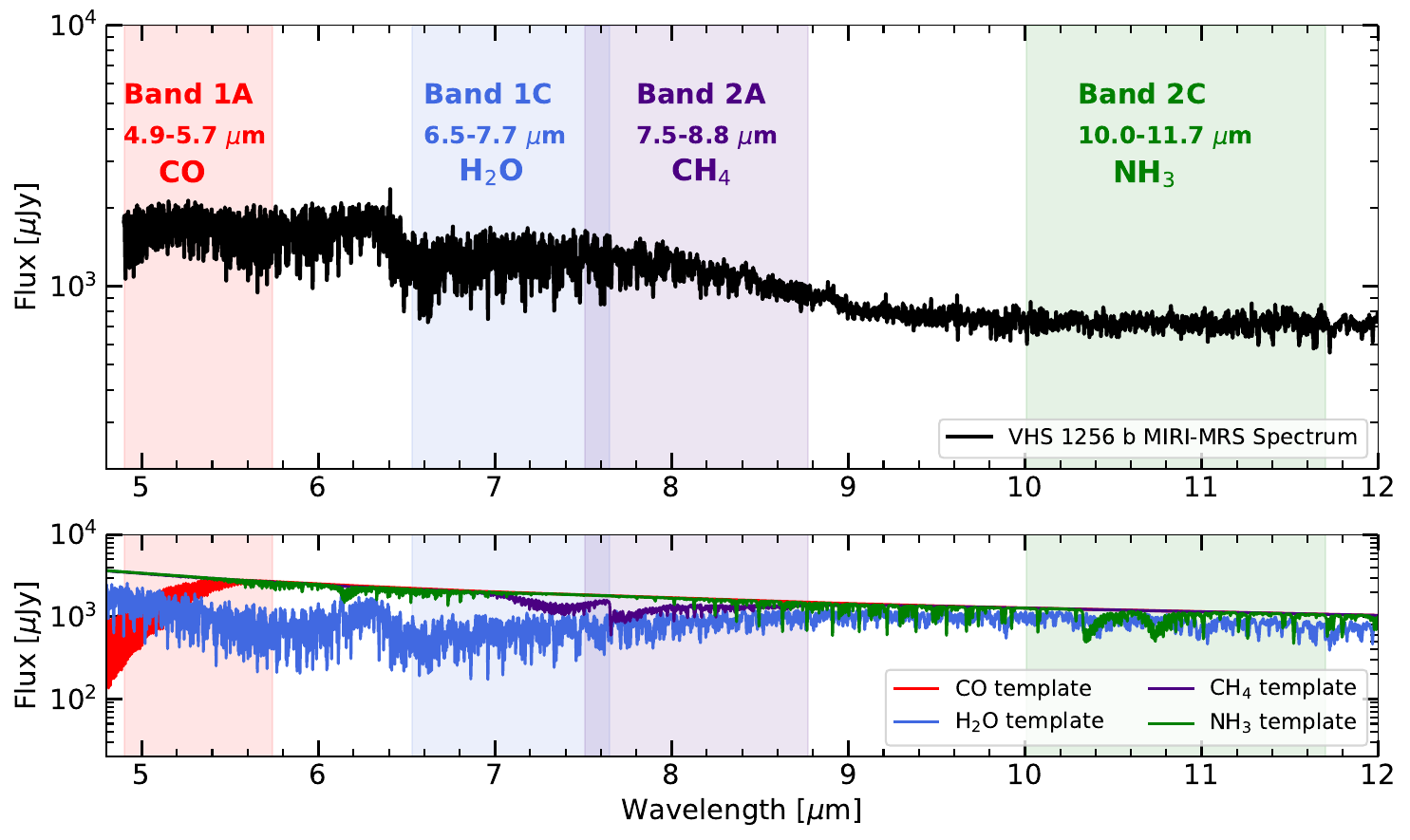}
    \caption{Top: MIRI-MRS spectra of VHS\,1256\,b, with the spectral regions used for molecular detection highlighted for each species. 
    Bottom: Corresponding \texttt{Exo-REM} molecular templates for the detected molecules, showing both the highlighted bands and the specific spectral features that were identified in the observations through cross-correlation.}
    \label{fig:spectrum_bands_templates}
\end{figure}

Fig.~\ref{fig:spectrum_bands_templates} shows the mid-IR spectrum of VHS~1256~b with the spectral bands used for the molecular detections highlighted. 
Several molecules are detected in multiple bands, as discussed in Sect.~\ref{sec:mm_application} and Sect.~\ref{sec:ccf_1d}. 
The bottom panel shows again the molecular templates, highlighting the same spectral regions.
We note that none of the molecular features are readily visible by eye (aside from the strongest H$_2$O features), 
which illustrates the importance of correlation analyses to confirm their presence.

\section{Molecular mapping}
\label{app:artifacts}

\begin{figure}[h]
    \centering
    \includegraphics[width=9cm]{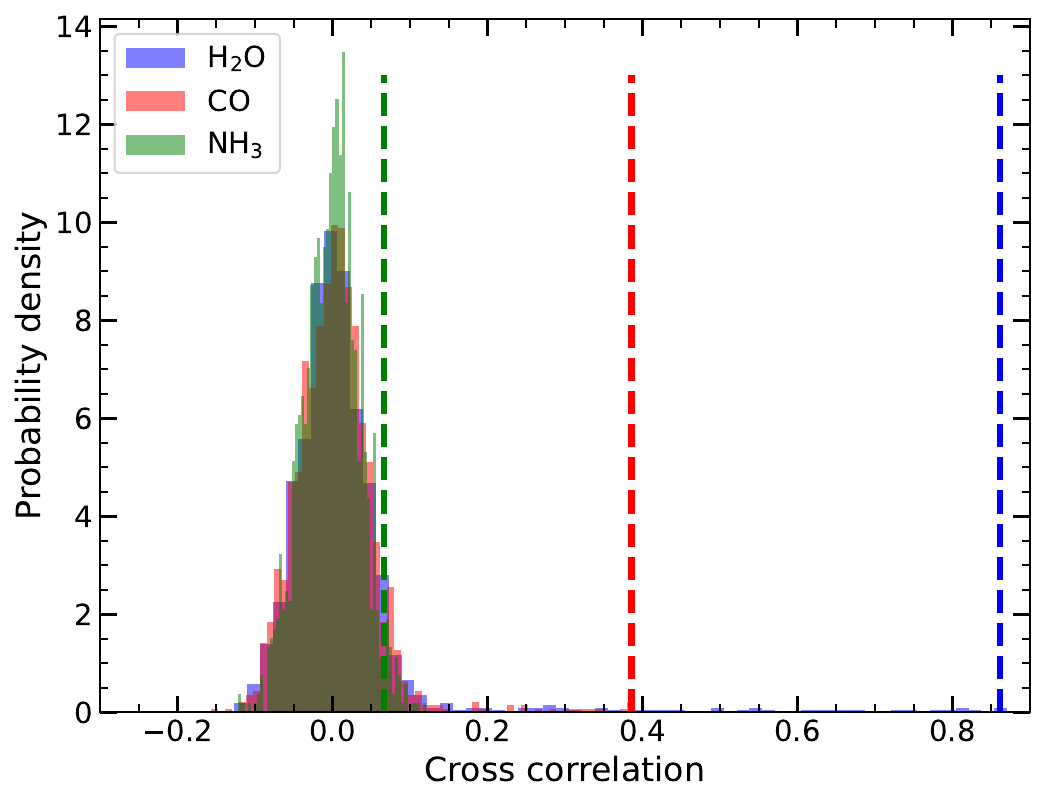}
    \caption{
    Histogram of the pixels in the correlation map (corresponding to the maps in Fig.~\ref{fig:corr_maps_molec}), 
    compared to the correlation coefficient at the location of the planet (dashed line).}
    \label{fig:histogram_cc_maps}
\end{figure}

This section provides additional figures supporting the molecular mapping analysis described in Sect. \ref{sec:mm_app_molecules}.
Figure \ref{fig:histogram_cc_maps} presents the distribution of the correlation values in the form of histograms. 
The noise follows here approximately a Gaussian distribution, with a dispersion consistent with the noise level estimated for the corresponding correlation maps. 
The correlation value measured at the planet’s location is indicated by the dashed vertical line in each panel and lies well outside the noise distribution in all cases.

\begin{figure}[h]
    \centering
    \includegraphics[width=9cm]{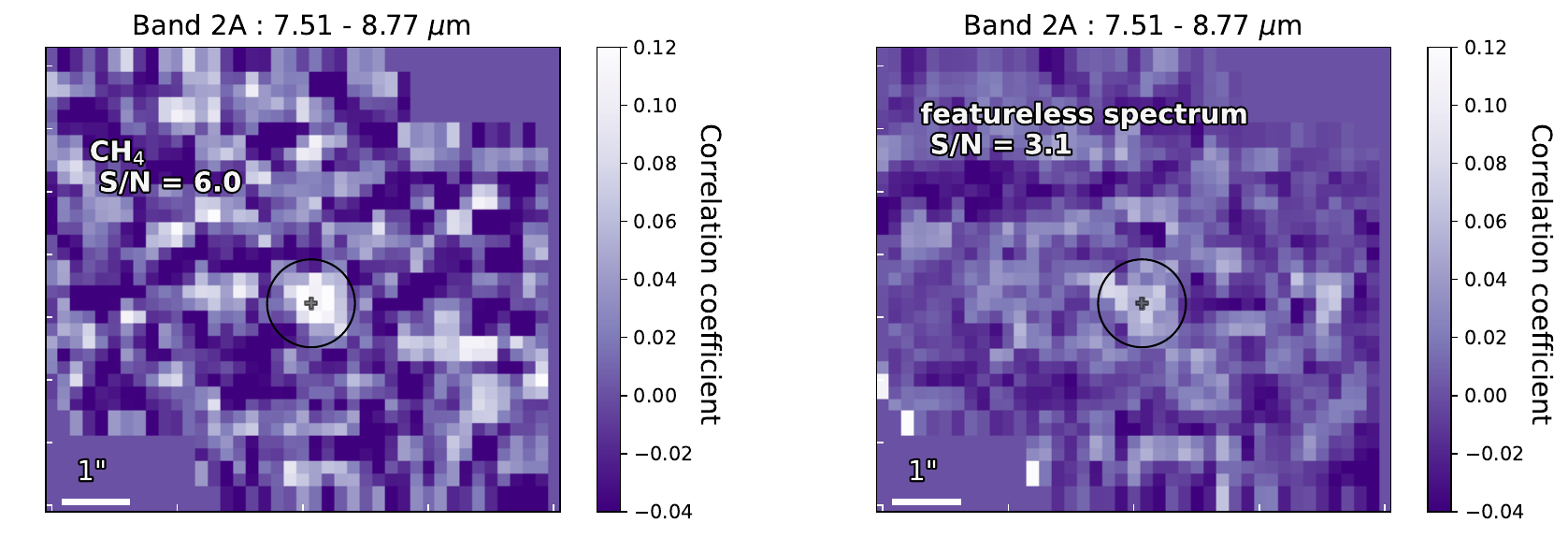}
    \includegraphics[width=9cm]{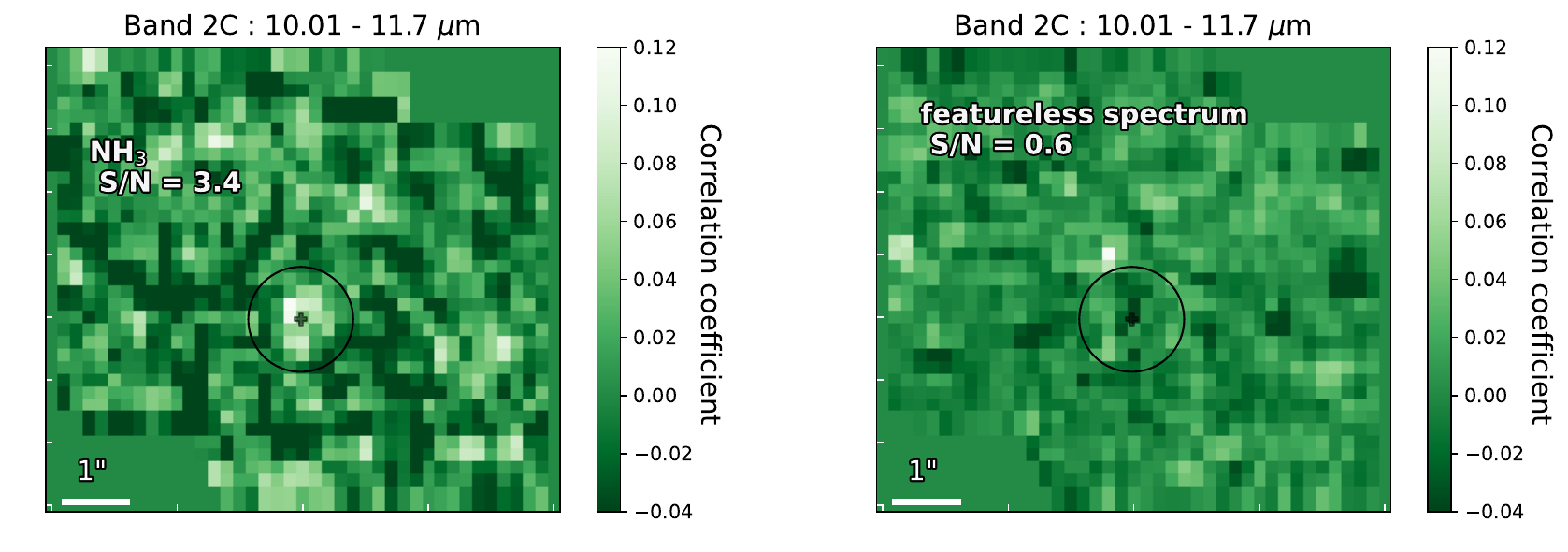}
    \caption{Left: Correlation maps with a CH$_4$ spectrum (purple, top) in band 2A, and with an NH$_3$ spectrum (green, bottom) in band 2C; corresponding to the spectral bands in which these molecules are expected to have absorption features in a planetary atmosphere.
    Right: Correlation maps with a featureless spectrum within the same spectral band.
    The same color scale is used for clear comparison.}
    \label{fig:artifacts}
\end{figure}

Figure~\ref{fig:artifacts} shows the correlation maps for CH$_4$ and NH$_3$ (left panels). 
The right panel presents the correlation map obtained from a featureless template spectrum (the CIA spectrum shown in Fig.~\ref{fig:ExoREM_molec_templates}), after continuum removal using Gaussian filtering.
This produces a “noise map”, illustrating the behavior of the correlation values when no molecular detection is expected.
Band 2A displays an overestimated $S/N$ ratio, which leads to an artificially enhanced CH$_4$ detection. This issue does not affect band 2C, where we obtain a robust NH$_3$ detection with $S/N = 3.4$.

\section{Correlation in each spectral band}
\begin{figure*}[h]
    \centering
    \includegraphics[width=18cm]{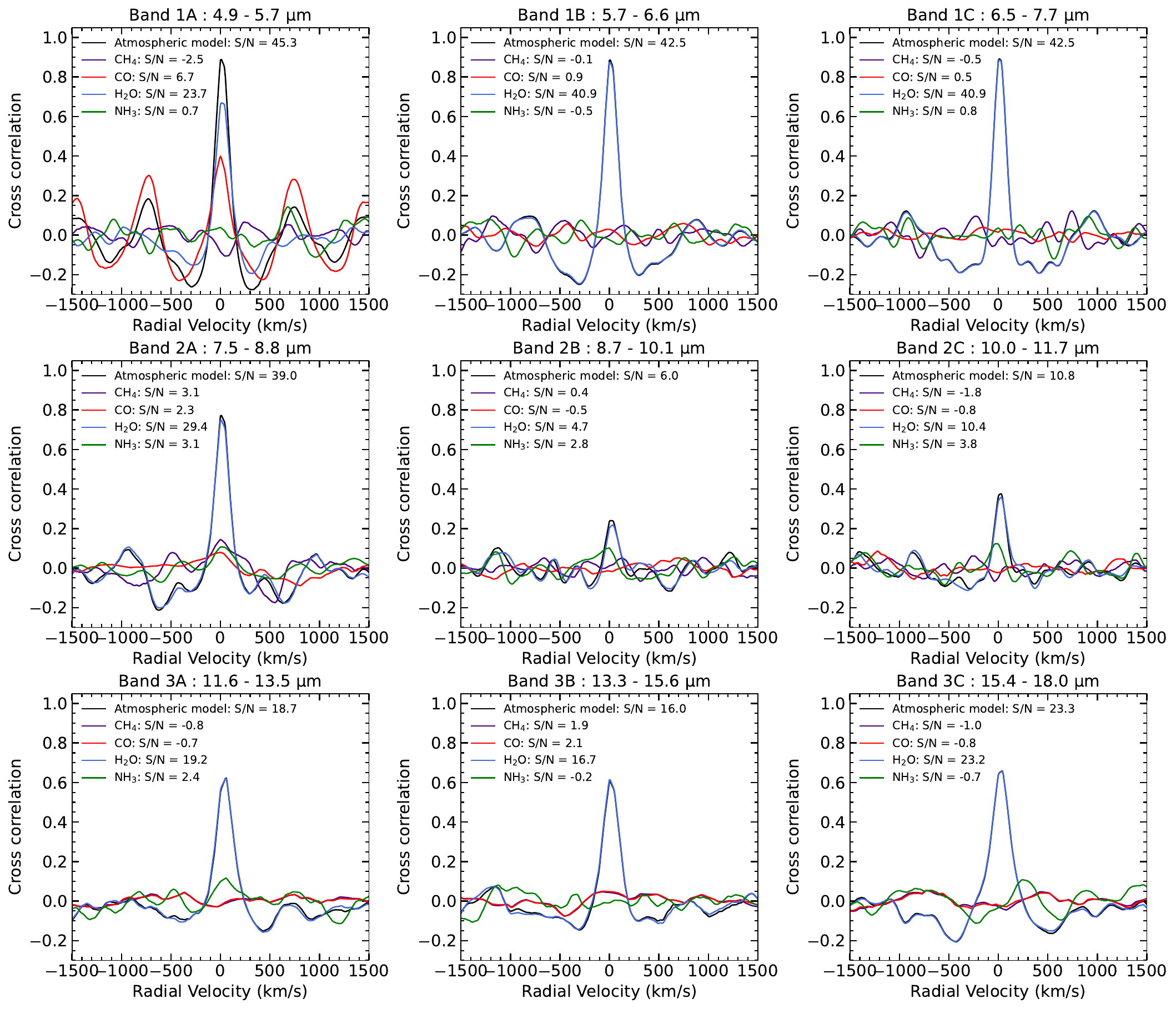}
    \caption{Cross-correlation functions of VHS\,1256\,b spectra with different molecular templates: full atmospheric model, H$_2$O, CO, NH$_3$ et CH$_4$ for all MIRI MRS bands.
    The $S/N$ is measured after removing the autocorrelation signal in the wings.}
    \label{fig:ccf_molec_all_bands}
\end{figure*}

\begin{figure}
    \centering
    \includegraphics[width=1\linewidth]{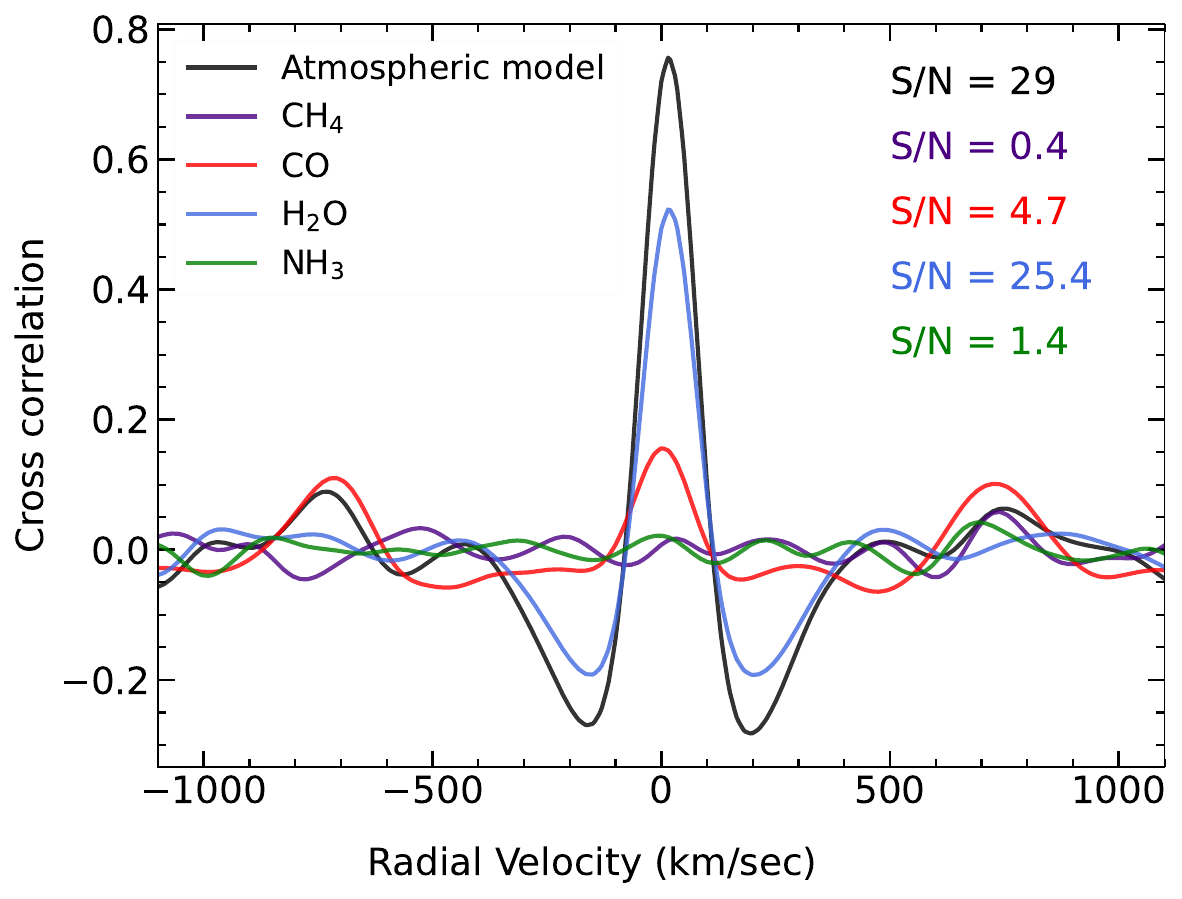}
    \caption{Cross-correlation measured across then entire MIRI-MRS wavelength range from 4.9 to 18 $\mu$m. The S/N values are indicated afer correcting for the autocorrelation.}
    \label{fig:CCFs_entire_wave_range}
\end{figure}

Supplementary figures for Sect.~\ref{sec:ccf_1d}.
Figure \ref{fig:ccf_molec_all_bands} presents the cross-correlation functions for each MIRI-MRS spectral band individually.
The detection is weaker in bands 2B and 2C, likely due to absorption by clouds. Molecules are most clearly detected in the spectral bands where their features are most abundant. 
Figure \ref{fig:CCFs_entire_wave_range} shows the same cross-correlation but taking into account direcly the entire spectral range from 4.9 to 18 $\mu$m. For CH$_4$ and NH$_3$, detections are stronger when focusing on the spectral bands containing the strongest molecular features, whereas considering a broader wavelength range leads to non-detections.
\label{app:ccf_1d}


\section{Comparison with forward-modeling results}
\begin{figure*}
    \centering
    \includegraphics[width=1\linewidth]{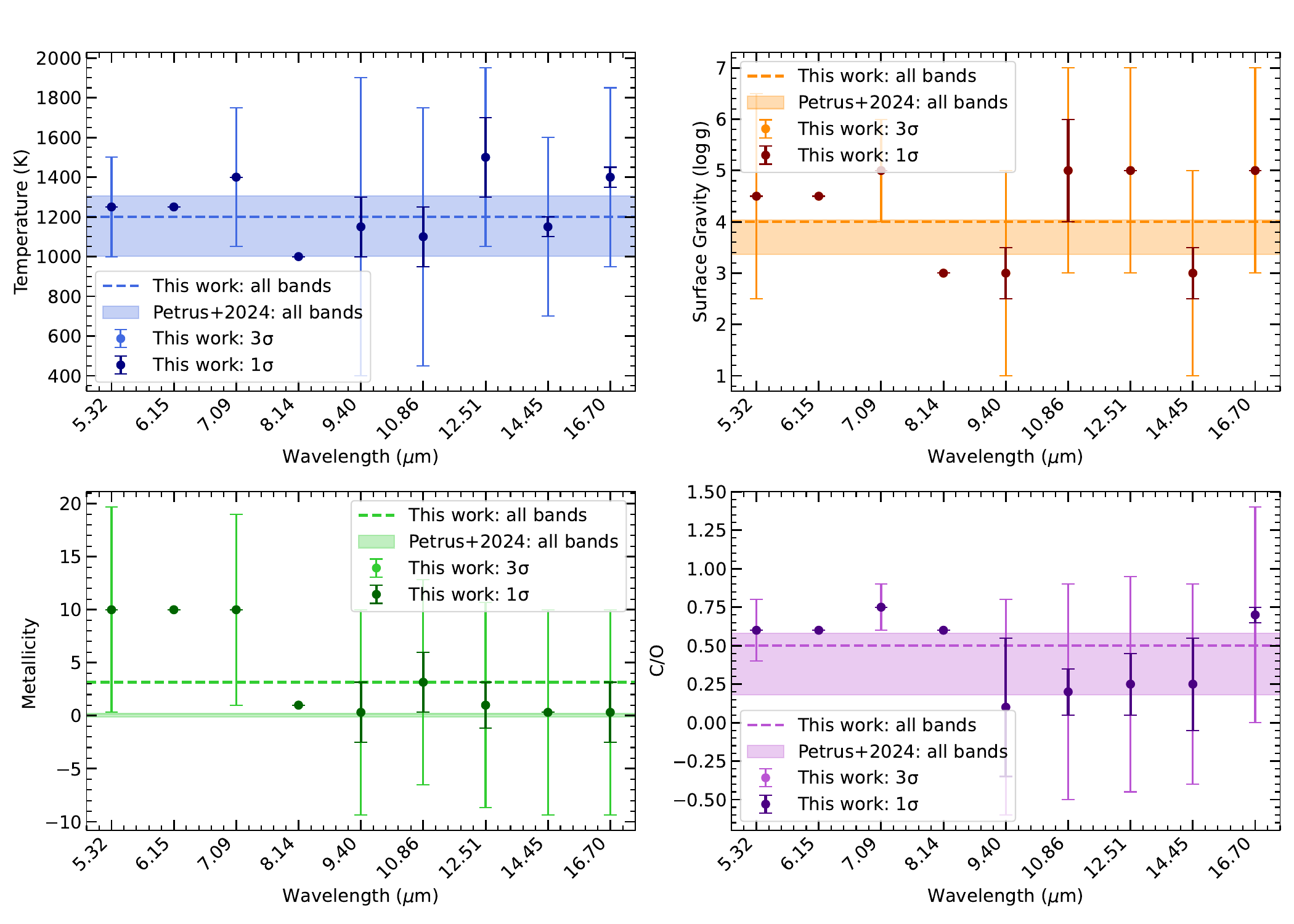}
    \caption{Measured temperature, surface gravity, metallicity, and C/O ratios derived from the correlation analysis in each MIRI-MRS spectral band. 
    Both uncertainties are 1-, and 3- $\sigma$ are indicated.
    The dashed lines correspond to the same measurements obtained using the full 5–18~$\mu$m MIRI spectrum.
    Metallicity is expressed relative to the solar value as $Z/Z_{\odot}$.
    The shaded areas indicate the corresponding measurements from \cite{petrus_jwst_2024} with the entire MRS spectral range and using the same \texttt{Exo-REM} amtopsheric model grid.}
    \label{fig:Comparison_correlation_formosa}
\end{figure*}

Figure~\ref{fig:Comparison_correlation_formosa} presents the atmospheric parameters of VHS\,1256\,b derived in our analysis (Sect.~\ref{sec:exorem_grid_ccf}) and compares them with the values reported by \cite{petrus_jwst_2024}, which were obtained using forward modeling of the full spectrum with the same \texttt{Exo-REM} grid.

\section{\texttt{Exo-REM} Template Spectra for NH$_3$ Abundances and CO Isotopic Ratios}
\begin{figure*}
    \centering
    \includegraphics[width=\linewidth]{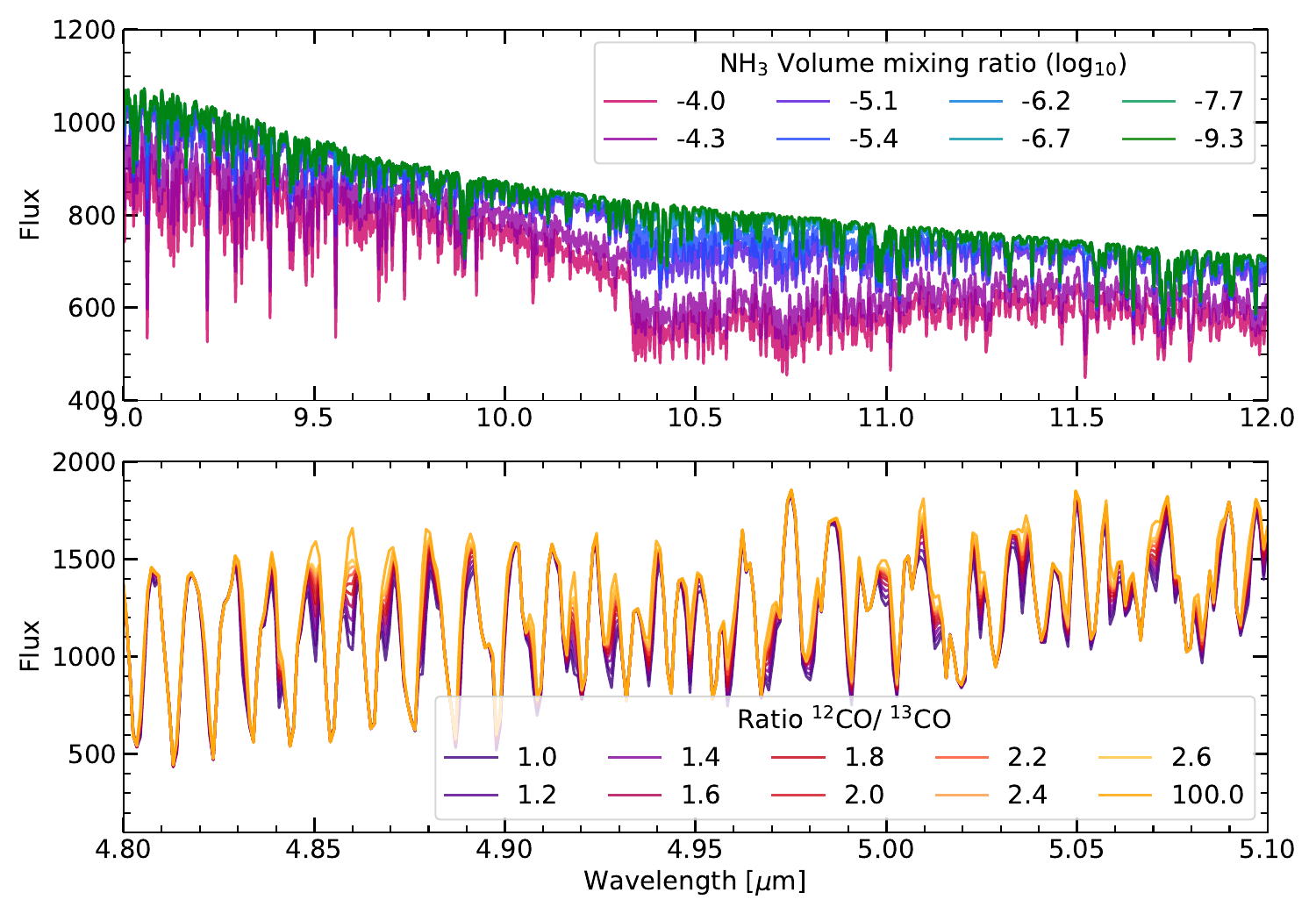}
    \caption{\texttt{Exo-REM} grids used in Sect.~\ref{sec:measuring_abundance}. 
    Top: Atmospheric models with various NH$_3$ abundances; only a subset of models from the grid is shown for clarity.
    Bottom: Atmospheric models with different CO isotopic ratios.}
    \label{fig:nh3_spectra}
\end{figure*}

Figure \ref{fig:nh3_spectra} shows the atmospheric model template spectra for various NH$_3$ volume mixing ratios, displaying a subset of the grid used in Section \ref{sec:measuring_abundance}.
The green spectrum corresponds to the lowest NH$_3$ abundance, while the pink spectra correspond to the highest. 
The bottom panel presents atmospheric models with different $^{12}$CO/$^{13}$CO isotopic ratios, ranging from 1 to 2.6. An additional spectrum with no $^{13}$CO ($^{12}$CO/$^{13}$CO = 100) is shown in orange for comparison.

\bibliography{references}{}
\bibliographystyle{aasjournalv7}

\end{document}